This is an Accepted Manuscript of an article published by Taylor & Francis in the International Journal of Human-Computer Interaction on 30 March, 2018, available online: http://www.tandfonline.com/10.1080/10447318.2018.1452351. For an eprint of the final published article, please access: https://www.tandfonline.com/eprint/T9d4cNwwRUqXPPiZYm8Z/full (limited number).

# Using Variable Dwell Time to Accelerate Gaze-based Web Browsing with Two-step Selection

Zhaokang Chen and Bertram E. Shi

The Hong Kong University of Science and Technology, Hong Kong

## Author Note

This work was supported in part by the Hong Kong Research Grants Council General Research Fund grant #16209014.

Zhaokang Chen, Department of Electronic and Computer Engineering, the Hong Kong University of Science and Technology, Clear Water Bay, Kowloon, Hong Kong

Bertram E. Shi, Department of Electronic and Computer Engineering and Department of Chemical and Biological Engineering, the Hong Kong University of Science and Technology, Clear Water Bay, Kowloon, Hong Kong.

Correspondence concerning this article should be addressed to Bertram E. Shi, Department of Electronic and Computer Engineering and Department of Chemical and Biological Engineering, the Hong Kong University of Science and Technology, Clear Water Bay, Kowloon, Hong Kong. E-mail: eebert@ust.hk

**ABSTRACT**

In order to avoid the "Midas Touch" problem, gaze-based interfaces for selection often introduce a dwell time: a fixed amount of time the user must fixate upon an object before it is selected. Past interfaces have used a uniform dwell time across all objects. Here, we propose a gaze-based browser using a two-step selection policy with variable dwell time. In the first step, a command, e.g. "back" or "select", is chosen from a menu using a dwell time that is constant across the different commands. In the second step, if the "select" command is chosen, the user selects a hyperlink using a dwell time that varies between different hyperlinks. We assign shorter dwell times to more likely hyperlinks and longer dwell times to less likely hyperlinks. In order to infer the likelihood each hyperlink will be selected, we have developed a probabilistic model of natural gaze behavior while surfing the web. We have evaluated a number of heuristic and probabilistic methods for varying the dwell times using both simulation and experiment. Our results demonstrate that varying dwell time improves the user experience in comparison with fixed dwell time, resulting in fewer errors and increased speed. While all of the methods for varying dwell time resulted in improved performance, the probabilistic models yielded much greater gains than the simple heuristics. The best performing model reduces error rate by 50% compared to 100ms uniform dwell time while maintaining a similar response time. It reduces response time by 60% compared to 300ms uniform dwell time while maintaining a similar error rate.

## INTRODUCTION

The use and capabilities of eye trackers have expanded rapidly in recent years. A state-of-the-art eye tracker can estimate eye gaze on a monitor screen accurately enough to allow users to interact with a computer system. This is especially useful for people with motor disabilities (Poole & Ball, 2006). Eye movement can be regarded as a very fast interaction modality and can be very informative about users' intent. As a result, use of eye movements is getting more and more attention in the field of human computer interaction (HCI) (Majaranta & Bulling, 2014).

Many innovative human computer interfaces have been designed using eye tracking devices. One of the most common uses of eye gaze is for object selection (Majaranta, Ahola, & Špakov, 2009; Zander, Gaertner, Kothe, & Vilimek, 2010). Enabling a user to select an object by looking at it might seem quite simple and intuitive. However, one of the most challenging problems for gaze interfaces is the "Midas Touch" problem: it is difficult to distinguish between the spontaneous eye movements used to gather visual information and the intentional eye movements for explicit selection (Jacob, 1995). The most common way to avoid unintentional selection is to introduce a dwell time: users must maintain their eye gaze on an object for a predefined duration before it is selected (Majaranta *et al.*, 2009; Murata, 2006; Räihä & Ovaska, 2012).

Some user interfaces avoid this problem by using gaze only to augment other input modalities. These "attentive user interfaces" (Majaranta & Bulling, 2014) only monitor the users' natural eye movements subtly in the background to infer their intent, and then modify the behavior of the primary user interface appropriately. Çığ and Sezgin (2015) used natural eye movements to improve real-time recognition of manipulation commands in a pen-based device using a hidden Markov model (HMM). Dong *et al.* developed a hybrid gaze/electroencephalography (EEG) interface which suppressed the selection of unlikely commands depending on the user's intent estimated from their natural gaze trajectories (Dong, Wang, Chen, & Shi, 2015; Wang, Dong, Chen, & Shi, 2015). In the field of human-robot interaction, Das, Rashed, Kobayashi, and Kuno (2015) designed a robot which was able to determine the suitable time to interact with a person by estimating the person's visual focus of attention based on the cues including gaze patterns. Huang and Mutlu (2016) used eye gaze in a robot system that delivered an item to a user who selected it from a menu using speech. By anticipating the eventual selection, the robot was able to plan or to perform anticipatory actions in advance, increasing task completion speed. Rozado, El Shoghri, and Jurdak (2015) used gaze trajectories improved the speed of web browsing using a standard interface by prefetching websites corresponding to hyperlinks upon which the users had fixated for more than 300ms. Similarly, Lee, Yoo, and Han (2015) reduced the initial delay in web video access by prefetching videos of potential interest.

This paper focuses on dwell-time based selection. In particular, we consider a two-step selection policy, where users first select a command by dwelling on a button at the side of the screen, and then select the object of the command by dwelling on that. Although there are other possible dwell based control paradigms, e.g., Lutteroth, Penkar, and Weber (2015), we focus on this two-step control policy because it is commonly used in practice and has been adopted in commercially available packages. For example, it is used by the Windows Control software package developed by Tobii Dynavox, which enables users to control their computer using only their eyes. It is also used in the GazeTheWeb browser (Menges, Kumar, Müller, & Sengupta, 2017).

Past work using dwell time, including those described above, has considered only the use of a fixed dwell time applied uniformly across all objects of potential interest. The choice of dwell time is a tradeoff between speed and the rate of unintentional selection: the shorter the dwell time, the faster objects can be selected, but the higher the rate of unintentional selection. Most systems allow users to adjust the dwell time to suit their personal taste.

In this paper, we propose a gaze-based browser where the dwell time required to select each hyperlink differs and changes dynamically over time depending upon the probability that the user wishes to select it, as estimated from the past gaze trajectory. This work follows a similar vein as Mott *et al*. (2017) and Pi & Shi (2017), who proposed dynamic dwell time approaches for gaze-typing. These systems dynamically adjust the dwell times of different letters based on the probabilities given by a letter prediction engine, resulting in faster and more accurate typing performance compared to the static uniform dwell time approach. The work reported here differs in that we consider a different application domain, gaze-based web-browsing. We also propose a different model for eye gaze, due to the differences in eye movement behavior during gaze typing and web browsing.

This paper makes four main contributions. First, we develop a new probabilistic model of eye gaze behavior during web surfing. This model enables us to estimate the probability that the user wishes to select each hyperlink on the web page based on his/her natural gaze behavior. Second, we propose a variable dwell time assignment policy, which sets the dwell time associated with each hyperlink depending upon its probability of being the target. Links that are more likely to be of interest have shorter dwell times to increase response speed. Less likely links have longer dwell times to avoid unintentional selection. Third, we provide the specific guidance in choosing the parameters of the dwell time assignment policy so as to achieve the best tradeoff between selection speed and false selection. Fourth, we quantitatively evaluate performance of the proposed method through both simulation and experiment, and compare it with simpler heuristics for varying the dwell time. Our results show that while heuristic methods for varying the dwell time can improve user experience, our probabilistic model gives much greater improvements in both speed and accuracy. Our best performing method reduces error rate by 50% while achieving similar response time compared to 100ms uniform dwell time. It reduces response time by 60% while achieving similar error rate compared to 300ms uniform dwell time. Because the basic paradigm for selection is unchanged, the use of variable dwell time does not increase cognitive load on the users.

## GAZE-BASED WEB BROWSER INTERFACE

The graphical user interface of the gaze-based web browser used in this study is presented in Figure 1. It is written in JavaScript as a Chrome Extension. It can communicate with MATLAB via a Chrome App using TCP/IP protocol. The user interface captures the locations and sizes of the bounding box of all hyperlinks on current page and sends them to MATLAB, which implements the probabilistic model described in next section.

This user interface provides a task bar on the right hand side of the web browser, which contains four command buttons, “Back”, “Select”, “Cancel” and “Forward”. The “Back” (“Forward”) button is used to go backward (forward) in the browsing history. The “Select” button is used to initiate the selection of a hyperlink to follow. The “Cancel” button is used to cancel the most recent command.

Users select the “Back”, “Cancel” and “Forward” commands using a single-step policy with constant dwell time. These commands are activated if the user maintains his/her gaze on the corresponding button for a fixed dwell time of 400ms. Once activated, the button color turns red, indicating to the user the identity of the selected command and the time it is selected. Users are not given instantaneous feedback about the system’s estimate of their current gaze location, as this may be distracting (Jacob, 1995).

The selection of hyperlinks follows a two-step policy. Users first select the “Select” button by gazing at it for 400ms. Then, they select the desired hyperlink by gazing at it for a dwell time that varies between hyperlinks. The hyperlink $m$ is selected if $N_m$ of the most recent $1.5N_m$ gaze points are assigned to that hyperlink. We define $T_m = N_m T_s$ to be the dwell time of hyperlink $m$, where $T_s = 16.67$ms is the sampling period of the gaze tracker. The larger window over which gaze points can be accumulated makes the system more tolerant of noise and jitter during fixation. Each gaze point, $G = (x_g, y_g)$, is assigned to the closest hyperlink according to the distance measure

$$D(G,m) = \min_{(x,y)\in B_m} \left\{ \max\left( |x_g - x|, |y_g - y| \right) \right\} \quad (1)$$

where $B_m$ is the bounding-box of hyperlink $m$, as long as this distance is smaller than or equal to 40 pixels. If the distance is larger than 40 pixels, then $G$ is not assigned to any hyperlink. Once a hyperlink is selected, its text color changes from blue to red.

## METHODS FOR VARYING DWELL TIME

Intuitively, the hyperlinks viewed immediately before the activation of the “Select” button are more likely to be the hyperlink that the user wishes to select, in comparison with hyperlinks that were not viewed or viewed in the more distant past. This has been observed in different domains. For example, when users were instructed to choose the more attractive of two faces, Shimojo, Simion, Shimojo, and Scheier (2003) found that gaze was biased towards the more attractive face, especially during the time immediately preceding the decision. To evaluate the extent to which this intuition holds for the case of gaze browsing, we analyzed eye gaze data from Experiment I described below, where users freely browsed the web through the gaze browser. Figure 2 shows that the selected hyperlink was among the last three hyperlinks fixated upon before “Select” was chosen in about 80% of the time. However, the selected hyperlink corresponded to the most recent fixation in only about 56% of the time. This suggests that the recent gaze history is informative, and that it makes sense to take into account more history than only the most recent fixation.

It seems reasonable to assign shorter dwell times to more recently viewed hyperlinks. However, there are a number of different ways in which one might assign different dwell times. It is not immediately obvious which would result in the best performance. In this section, we describe the methods we studied for varying the dwell time: three heuristic methods, and a data-driven probabilistic method. All methods depend upon a first stage that segments the gaze trajectory into a sequence of fixations.

**Fixation Segmentation**

This section describes the method we used to segment the raw eye gaze trajectory into the scanpath (Noton & Stark, 1971): a sequence of fixations separated by saccades.

We model reading behavior as a sequence of fixation-saccade cycles. Accurate identification of fixations is important because visual information is obtained during fixations (Rayner, 1998). The location and duration of fixations usually indicate the points of interest to the person. Thus, we propose an algorithm which automatically identifies fixations from sampled gaze data from eye tracker.

In contrast to algorithms which require choosing threshold parameters manually, e.g., velocity-threshold identification (I-VT) and dispersion-threshold identification (I-DT) (Salvucci & Goldberg, 2000), we use here a data driven algorithm. We extend the hidden Markov model identification (I-HMM) (Salvucci & Goldberg, 2000), which utilized a first-order autoregressive first-order HMM to label each gaze point as fixation or saccade, to a second-order autoregressive second-order hidden Markov model. The I-HMM was vulnerable to noise because it only considered first-order dependencies. Introducing a higher order dependency makes our model more robust to variation and noise, and enables us to address the problem of single outliers during fixation (M. Kumar, 2007). We denote our model as I-HMM2.

The I-HMM2 labels each gaze point as a fixation, saccade or outlier. The evolution of joint probability of sequence of hidden states (labels, $l_1, l_2, \ldots l_T$) and sequence of observations (gaze points, $g_1, g_2, \ldots g_T$) is given by:

$$p(l_1,l_2,\cdots l_T,g_1,g_2,\cdots g_T)=p(l_1,l_2,g_1,g_2)\prod_{t=3}^{T}p(l_t\mid l_{t-2},l_{t-1})\prod_{t=3}^{T}p(g_t\mid l_{t-2},l_{t-1},l_t,g_{t-2},g_{t-1}) \quad (2)$$

where $l_t \in$ {fixation (*f*), saccade (*s*), outlier (*o*)} is the hidden label of $g_t$. Transitions between state *o* and state *s* are not allowed, as we assume that the outliers only occur during fixations. Self-transitions within state *o* are not allowed either for the same reason.

The emission distributions, $p(g_t|l_{t-2},l_{t-1},l_t,g_{t-2},g_{t-1})$, are assumed to be two-dimensional Gaussian distributions, i.e., $g_t \sim N(\mu_e(l_{t-2},l_{t-1},g_{t-2},g_{t-1}),\Sigma_e(l_{t-2},l_{t-1},l_t))$, where $\mu_e(l_{t-2},l_{t-1},g_{t-2},g_{t-1})$ is given by Table 1 and $\Sigma_e(l_{t-2},l_{t-1},l_t)$ is given by Table 2 if $l_t = f$, $\Sigma_s$ if $l_t = s$ or $\Sigma_o$ if $l_t = o$, respectively. The covariance matrices $\Sigma_f$, $\Sigma_s$ and $\Sigma_o$ are assumed to be diagonal and learned from data. The design of $\mu_e$ and $\Sigma_e$ was based on the assumption that the gaze points belonging to one fixation are independent and identically distributed (*i.i.d.*) variables following $N(\mu_f, \Sigma_f)$, where $\mu_f$ is an unknown fixation location. According to this assumption, if $g_{t-2}$, $g_{t-1}$ and $g_t$ belong to the same fixation, then $g_t - g_{t-2} \sim N(0,2\Sigma_f)$, $g_t - g_{t-1} \sim N(0,2\Sigma_f)$ and $g_t-(g_{t-2}+g_{t-1})/2 \sim N(0,1.5\Sigma_f)$.

The initial probability, $p(l_1,l_2,g_1,g_2)$ is assumed to be uniformly distributed over hidden states $(l_1,l_2)$ and screen coordinates $(g_1,g_2)$. This setting has little impact on the performance of the model as it runs at 60Hz (the sampling rate of the gaze tracker) with long sequence of data.

Given a sequence of gaze points, $g_1,g_2,\ldots g_T$, we use the Viterbi algorithm (Rabiner, 1989) to compute optimal label sequence. The parameters of this first stage model can be trained from unlabeled data by modifying the Baum-Welch algorithm (Rabiner, 1989). Note that only the number of parameters of transition matrix increases (from six to 14) when extending the first-order model to a second-order model.

We use the label sequence to convert the raw eye gaze data into sequence of fixations. We first group the gaze data between each pair of neighboring saccades into a single fixation. The fixation location $(x,y)$ is obtained by averaging the locations labeled as fixations. The duration $d$ is is defined as the time between the two saccade boundaries. Fixations with durations less than 100ms are discarded, since fixations are rarely shorter than 100ms (Salvucci & Goldberg, 2000).

**Heuristic Methods for Assigning Dwell Times**

We propose three heuristic methods to adjust the hyperlink dwell times according to the scanpath. These methods first assign a partial ranking to the hyperlinks on the page. They then assign shorter dwell times to the first $k$ hyperlinks, where higher ranked hyperlinks have shorter dwell times, and a single longer dwell time to all remaining hyperlinks.

We considered three different ranking schemes: distance-based, temporal and duration-based. Distance-based ranking ranks the hyperlinks according to their distance from the last fixation point, as measured according to (1). Closer hyperlinks are given higher rankings. Temporal ranking associates each fixation with the hyperlink closest to it according to (1), and then ranks hyperlinks according to the order of fixations. The hyperlink associated with the most recent fixation has the highest rank. Duration-based ranking also associates each fixation with its closest hyperlink, but ranks them in descending order of fixation duration.

To assign dwell times based on the ranking, we assigned all hyperlinks outside the top-$q$ to have the longest dwell time $\overline{T}$. The dwell times of the top-$q$ candidates decreased linearly to the shortest dwell time $\underline{T}$. We define the parameters of the heuristic policies to be $[\overline{T}, \underline{T}, q]$.

**Probabilistic Method for Assigning Dwell Times**

The probabilistic method uses a probabilistic model of the scanpath to infer the likelihood that each hyperlink on the page corresponds to the target. It then assigns shorter dwell times to more likely hyperlinks.

We use a factorial hidden Markov model (FHMM) to model the scanpath before the select action is chosen. The FHMM generalizes the HMM by assuming that the observations depend upon multiple latent variables which evolve according to independent Markov chains. The independence assumption enables the combined state transition matrix to be factorized, reducing the number of parameters (Ghahramani & Jordan, 1997).

Figure 3 presents the FHMM model as a Bayesian network. The two latent variables are the intended target link for selection, $I_t$, and the gaze behavior, $B_t$. The intended target $I_t$ assumes integer values from 1 to $M$, the number of hyperlinks on the current page. The gaze behavior, $B_t$, assumes one of three integer values: 1 (reading the hyperlink), 2 (reading near the hyperlink), and 3 (reading away from the hyperlink). Both latent variables evolve according to independent Markov chains which end at a terminal state (TS), corresponding to the activation of the "Select" button. The location and duration of the $t^{\text{th}}$ fixation in the sequence, $f_t$, depends on both $I_t$ and $B_t$.

The joint probability of hidden states, $z_t = [I_t\ B_t]$, and observations, $f_t$, can be calculated by:

$$p(z_1, z_2, \cdots z_T, f_1, f_2, \cdots f_T) = p(z_1)p(f_1 \mid z_1)\prod_{t=2}^{T} p(z_t \mid z_{t-1})\prod_{t=2}^{T} p(f_t \mid z_t) \tag{3}$$

We assume that the combined state transition probabilities factorize according to

$$p(z_t \mid z_{t-1}) = p(I_t \mid I_{t-1})p(B_t \mid B_{t-1}) \tag{4}$$

We assume that the intended target transition probabilities have the form

$$p(I_t = k \mid I_{t-1} = j) = \begin{cases} 1 - p_s & , k = j \\ p_s/(M-1) & , k \neq j \end{cases} \tag{5}$$

where $p_s$, the probability that the target changes, is learned from data. The transition probabilities for gaze behavior are learned without constraints.

The spatial location $(x,y)$ and the duration $d$ of the fixation are assumed to be conditionally independent. Thus, the emission distribution, $p(f_t|I_t, B_t)$, factorizes as

$$p(f_t = [x, y, d] \mid I_t, B_t) \equiv p(x, y \mid I_t, B_t) \times p(d \mid B_t) \tag{6}$$

where $p(x, y \mid I_t = m, B_t = k)$ is given by a normal distribution with mean $\mu_m$ and covariance matrix $\Sigma_{m,k}$ when $k = 1$ or 2 and is given by a uniform distribution over the entire screen when $k = 3$. The mean $\mu_m$ is given by the center of the bounding box of hyperlink $m$. The covariance matrix is given by

$$\Sigma_{m,k} = \begin{bmatrix} (\beta_{x,k} W_m)^2 + \sigma_{x,k}^2 & 0 \\ 0 & (\beta_{y,k} H_m)^2 + \sigma_{y,k}^2 \end{bmatrix} \tag{7}$$

where $W_m$ and $H_m$ are the width and height of the hyperlink, respectively. The $\beta$ and $\sigma$ parameters are learned, and are constrained so that they are larger for $k = 2$ than $k = 1$. The distribution of the duration is assumed to be lognormal with mean and variance parameters that vary according to the behavior.

The initial probability distribution is given by

$$p(z_1) \equiv p(I_1 = m, B_1 = k) = \frac{\pi_k}{M} \tag{8}$$

where $\pi_k$ are the initial parameters to be learned.

Once the "Select" command is activated, we compute the probability that hyperlink $m$ is the intended target according to

$$p(I_T = m \mid z_{T+1} = \mathrm{TS}, f_1, \cdots f_T) = \frac{\sum_k p(I_T = m, B_T = k, f_1, \cdots f_T)\, p(z_{T+1} = \mathrm{TS} \mid B_T = k)}{\sum_{n,k} p(I_T = n, B_T = k, f_1, \cdots f_T)\, p(z_{T+1} = \mathrm{TS} \mid B_T = k)} \tag{9}$$

The probability $p(I_T = m, B_T = k, f_1 \cdots f_T)$ can be computed using the forward algorithm (Rabiner, 1989). We assume that the transition to the terminal state depends only on the gaze behavior and not the target identity, i.e.

$$p(z_{T+1} = \mathrm{TS} \mid z_T) = p(z_{T+1} = \mathrm{TS} \mid B_T) \tag{10}$$

We train the model parameters from data in a two-step process. We first train the gaze behavior transition probabilities and the parameters of the emission distributions in a semi-supervised manner. We use a slightly modified Baum-Welch algorithm, where we assume that the intended target $I_t$ is always equal to the known target for all $t$. We then fix these parameters, and find the optimal value of the $p_s$ parameter by grid-search in a second pass over the training data during which $I_t$ is free to transition.

This FHMM model accounts for variability in the locations, sizes and number of hyperlinks on different web pages. It also allows for variability in the length of the scanpath, and accounts for variability in the position of gaze due to noise or jitter through the use of the Gaussian distribution.

We assign different dwell times to each hyperlink using a dwell time assignment policy that is a non-increasing function of the probability that the hyperlink is the intended target. The nominal dwell time of hyperlink $m$ is given by

$$T_m = h\big(p(I_T = m \mid z_{T+1} = \mathrm{TS}, f_1 \cdots f_T)\big) \tag{11}$$

where we choose $h(\cdot)$ among a set of piecewise linear policies determined by four parameters [$T_{\mathrm{max}}$, $T_{\mathrm{min}}$, $T_{\mathrm{break}}$, $p_{\mathrm{break}}$],

$$h(p)=\begin{cases} T_{\max}-(T_{\max}-T_{\text{break}})\frac{p}{p_{\text{break}}} & \text{if } 0 \le p \le p_{\text{break}} \\ T_{\text{break}}-(T_{\text{break}}-T_{\min})\frac{p-p_{\text{break}}}{1-p_{\text{break}}} & \text{if } p_{\text{break}} < p \le 1 \end{cases} \tag{12}$$

Figure 4 shows two example policies. The function drops linearly from $T_{\max}$ at $p = 0$ to $T_{\text{break}}$ at $p = p_{break}$ and then to $T_{\min}$ at $p = 1$. To ensure the function is non-increasing, we constrain $0 < T_{\min} \le T_{\text{break}} \le T_{\max}$. Due to sampling by the gaze tracker, the actual dwell time is obtained by quantizing nominal dwell time to an integer number of samples.

# EXPERIMENTAL PROCEDURES

## Setup

During experiments, subjects were seated about 65cm in front of a 19 inch monitor with 1280×1024 pixel resolution. A chin-rest was used to keep their head position stable. Eye gaze data were recorded at 60Hz by a Tobii X60 eye tracker mounted under the monitor. Before each experiment, the eye tracker was calibrated using the standard nine-point calibration, followed by a validation session to check whether the eye tracker was accurate enough to distinguish between two adjacent rows on the web pages. If not, calibration was repeated. Calibration was also repeated during breaks if necessary.

Before each experiment, the subject was given a practice session to familiarize themselves with the gaze-based browser. The dwell times for all hyperlinks were fixed at 500ms for the practice session.

All of the web pages used in this study were taken from Wikipedia in English, as each page contains a large and diverse range of hyperlinks.

## Participants

In total, 34 subjects (21 males and 13 females) participated in this study. All had normal or corrected-to-normal vision. None had prior experience with gaze-based web browsing. We conducted four experiments, but as shown in Table 3, not all subjects participated in all four experiments. The experimental procedures involving human subjects described in this paper were approved by Committee on Research Practices at the Hong Kong University of Science and Technology. All subjects provided written consent.

**Experiment I**

In this experiment, subjects were free to browse the web using the gaze browser. The dwell times of hyperlinks were all set to 500ms.

Fourteen subjects (nine males and five females) participated in this experiment. Each subject performed one or two sessions depending on their self-reported fatigue level, with a break between each session. Each session of the experiment starting by presenting the user with a webpage of his/her choice, and then allowing him/her to choose and follow hyperlinks freely. The session ended after the user had followed five to six hyperlinks.

If a link was incorrectly selected, users were instructed to either cancel the selection and try again or report the misselection and the intended target link to the experimenter. This instruction allowed us to obtain the intended target links of the users.

Each session was divided into trials, where each trial spans the time from when a web page is first presented until the first attempt at selecting a hyperlink is made, irrespective of whether that attempt resulted in a correct or an incorrect selection by the system. On average, one trial lasted for about 30 seconds and contained 78 fixations.

In total, we collected gaze data from 86 valid trials. Five trials were considered invalid because of significant eye tracking inaccuracy. The gaze data from this experiment was used to train the gaze models used in all subsequent experiments.

**Experiment II**

While free viewing is a more natural task, its drawback is that it results in relatively few selections per unit time, as subjects spend most of their time reading the text of the webpage. To accelerate data collection and provide for better experimental control, subjects in Experiment II were asked to perform a search-and-select task, where they were asked to find and select particular hyperlinks. We collected data under four hyperlink dwell time assignment policies:

- Uniform: 500ms
- Uniform 300ms
- Uniform 100ms
- Probabilistic with parameters [600ms, 100ms, 100ms, 1]

For the probabilistic policy, the hyperlink dwell time decreased linearly from 600ms to 100ms with the inferred target probability. The actual dwell time was quantized to a multiple of 100ms. In this and all subsequent experiments, we used only the most recent $T = 5$ fixations of the scanpath to assign target probabilities to all hyperlinks. A preliminary study suggested that this saved computation time without too much sacrifice in accuracy (data not shown).

Twelve subjects (nine males and three females) participated in this experiment. Each subject participated in four sessions. Subjects were given a break between the second and third session. The sessions differed according to the dwell time assignment policy used and the set of web pages/desired hyperlinks presented to the users. All subjects used each policy once, but in random order. Subjects were told that the policies would vary between sessions, but were not told which policy they were using in each session.

Each session contained 10 trials. In each trial, the subjects were asked to find and select a particular hyperlink. The text of the desired hyperlink was presented to the user at the beginning of each trial, and also presented at the top of the screen during the trial in case the subjects forgot. A trial ended when a hyperlink was selected, regardless of the correctness. A trial was considered valid if the selection was made within 45s from the start of the trial and if a hyperlink was selected within two seconds after the “Select” button was activated.

We pre-defined four sets of web pages/desired hyperlinks to balance the difficulty among different sessions and among different subjects for better experimental control. Desired hyperlinks were chosen from more cluttered locations on the webpages. Each session used a different set, but the order in which the sets were presented was randomized independently of the policy order.

After each session, we asked the subjects to evaluate accuracy/response speed of the different policies using a 5-point Likert-like scale, where 1 means very inaccurate/slow and 5 means very accurate/fast. We also asked subjects to rank the policies based on their overall preference. The subjects could revise their scores/ranking at the end of experiment to ensure consistency.

In total, we collected data from 477 valid trials. One trial using the 100ms dwell time policy and two trials using the 300ms dwell time policy were discarded as invalid because the subjects failed to find the instructed hyperlink.

**Experiment III**

Experiment III was similar to Experiment II, except that we collected data from six different dwell time policies over six sessions:

- Uniform: 500ms
- Probabilistic with parameters [500ms, 16.67ms, 50ms, 0.005]
- Probabilistic with parameters [500ms, 16.67ms, 50ms, 0.3]
- Probabilistic with parameters [500ms, 16.67ms, 16.67ms, 1]
- Probabilistic with parameters [500ms, 100ms, 100ms, 1]
- Probabilistic with parameters [500ms, 316.67ms, 316.67ms, 1]

We chose these parameters based on the results of Experiments I and II to achieve close to the best tradeoff between speed and accuracy. The actual dwell time was quantized to a multiple of the eye tracker sampling period.

Ten subjects (seven males and three females) participated in this experiment. We collected data from 598 valid trials. One trial under the [500ms, 16.67ms, 50ms, 0.3] policy and one trial under the [500ms, 16.67ms, 16.67ms, 1] policy were invalid because the subjects failed to find the instructed hyperlink. Subjects were not asked to evaluate the different policies.

**Experiment IV**

Experiment IV was similar to Experiment II and III. We collected data from six different dwell time policies over six sessions:

- Uniform: 183.33ms
- Heuristic distance-based ranking with parameters [500ms, 83.33ms, 5]
- Heuristic temporal ranking with parameters [500ms, 100ms, 5]
- Heuristic duration-based ranking with parameters [500ms, 100ms, 5]
- Probabilistic with parameters [500ms, 16.67ms, 50ms, 0.9] and I-HMM-based fixation segmentation
- Probabilistic with parameters [500ms, 33.33ms, 33.33ms, 1] and I-HMM2-based fixation segmentation

We chose these parameters because their average response times are all expected to be approximately 300ms, based on simulations using the data of Experiment I, II and III. To ensure consistency with the probabilistic method, where we computed hyperlink target probabilities only using the data from the last five fixations, we set $q$=5 for the heuristic methods.

Similar to Experiment II, after each session, we asked the subjects to evaluate accuracy/response speed of the different policies using a 10-point Likert-like scale. We also asked subjects to rank the policies based on their overall preference.

Twelve subjects (seven males and five females) participated in this experiment. We collected data from 715 valid trials. One trial under distance-based ranking, one trial under duration-based ranking, one trial under the probabilistic method with I-HMM-based fixation segmentation and two trials under the probabilistic method with I-HMM2-based fixation segmentation were invalid because the subjects failed to find the desired hyperlink.

**Simulation Experiments**

We used the data collected in Experiments I, II and III when users were using the uniform 500ms dwell time policy to simulate the performance of different dwell time policies. This enabled us to predict the performance of a wide range of different dwell time assignment policies without extensive experimentation.

We performed our simulations by using the gaze trajectory before the user activated the “Select” button to either rank the hyperlinks when using the heuristic methods or to estimate the probability that each link was the target link when using the probabilistic model. We then used the ranking or probabilities to assign different dwell times to different hyperlinks. Finally, we predicted which hyperlink would be selected based on the gaze trajectory after the user activated the “Select” button.

We used grid search to perform exhaustive simulation over the entire range of dwell time assignment policies such that $0 < T_{min} \leq T_{max} \leq 500$ms. We varied $T_{max}$ from 16.66ms to 500ms, $T_{min}$ from 16.66ms to $T_{max}$, $T_{break}$ from $T_{min}$ to $T_{max}$, and $p_{break}$ from 0 to 1. The grid step size was 16.67ms (1 sample) in time and 0.1 in probability. Note that this search space includes the set of all uniform dwell time policies, which are obtained by setting $T_{\min} = T_{\max}$.

Since the maximum dwell time $T_{\max}$ used in simulation was not larger than the 500ms used in the experimental data and visual feedback was provided only after a hyperlink was selected, we expected that the dwell time assignment policy would not significantly influence gaze behavior. In the results section, we compare the results from simulation and experiment.

For each policy, we evaluated the average error rate and average response time. The error rate is defined as the percentage of incorrect hyperlink selections. The response time is defined as the duration between the moment when the user moves his/her eyes away from the "Select" button after activating it and the moment when a hyperlink is selected.

## RESULTS

### Target Inference from Gaze Trajectories

We used the data from Experiment I to train the eye gaze model parameters using fourteen-fold leave-one-subject-out cross-validation.

Table 4 and Table 5 show the learned parameter values for the FHMM scanpath model averaged over the fourteen folds, along with their standard deviations. The small values of the standard deviation indicate that parameters values varied little between folds. Figure 5 shows the emission distributions for the fixation location and duration.

We evaluated the model by testing its classification accuracy, which we defined as the percentage of the time the hyperlink *m* with the highest $p(I_T = m \mid z_{T+1} = \mathrm{TS}, f_1 \cdots f_T)$ was the same as the desired hyperlink. The average classification accuracy across the fourteen folds was 65.1%, much higher than the chance accuracy of 1.88% (Web pages in the testing set contained 53.30 hyperlinks on average). The classification accuracy is also higher than that obtained by the simple heuristic of assuming the last hyperlink fixated upon by the user is the desired (55.8%, Figure 2).

For a more nuanced characterization of the model, Figure 6 shows histograms of the probabilities $p(I_T = m \mid z_{T+1} = \mathrm{TS}, f_1 \cdots f_T)$ assigned to the true target link, the highest probabilities among the non-target links, and the differences between them. About 50% of the probabilities of the true target link were greater than 0.8, whereas over 50% of the probabilities of the most likely non-target link were smaller than 0.2. Although the model assigned a probability less than 0.2 to the true target about 30% of the time, it assigned a probability greater than 0.8 to the most likely non-target link only about 10% of the time. This suggests that when the true target link is assigned a low probability, most of the other hyperlinks also have low probabilities, i.e., the model is unsure.

This is consistent with the histogram of the differences: about 35% of the differences were greater than 0.8, whereas only about 10% of them were lower than -0.8.

Figure 7 shows the top-*k* accuracy of the rankings obtained from three heuristic models and the probabilistic method. We defined the top-*k* accuracy as the percentage of the time the desired hyperlink was one of the *k* highest ranked hyperlinks, where *k* varied from 1 to 5. The probabilistic model achieves the highest accuracy for all values of *k*. The probabilistic model with I-HMM2-based fixation segmentation has a top-1 accuracy that is 6.98% higher than the best performing heuristic model for the data from Experiment I, and 9.55% higher for the data from Experiments II, III and IV. To evaluate the effect of the fixation segmentation model, we also trained a probabilistic model using the I-HMM-based fixation segmentation algorithm. Using I-HMM2-based segmentation improves top-1 accuracy by about 3.2%, and maintains similar top-2 to top-5 accuracy. This suggests that I-HMM2 is a better model of gaze, resulting in better fixation segmentation.

**Simulation Results**

Figure 8 presents the results of simulations and experiments on probabilistic model with different dwell time assignment policies. Experimental results will be discussed in more detail in next section. For the simulation results, the performance of each policy is shown as a cyan marker. The commonly used uniform dwell time policy is shown as a red dash-dotted curve. The curve decreases, clearly illustrating the tradeoff between accuracy and speed. The lower right hand end corresponds to the 500ms policy. The upper left hand end corresponds to the 16.66ms policy (point not shown, response time: 31.10ms and error rate: 83.01%). Note that most of the cyan markers lie below and to the left of this curve, indicating that decreasing the dwell time based on the inference from the probabilistic gaze model leads to a lower error rate and/or a faster response time.

Through trial and error, we found two single parameter policies that approximated the lower left hand boundary of the region covered by the cyan dots, which represents the best tradeoff achievable by the class of non-decreasing piecewise linear policies considered. These policies are

I) [500ms, 16.67ms, 50ms, $p_{\text{break}}$], $p_{\text{break}} \in [0, 0.93]$

II) [500ms, $T_{\min}$, $T_{\min}$, 1], $T_{\min} \in [16.67\text{ms}, 500\text{ms}]$

Example policies are presented in Figure 9. The simulated performance of the two policies are shown as pink solid (Policy I) and blue dashed (Policy II) lines in Figure 8. The upper left ends of the lines correspond to $p_{\text{break}} = 0$ and $T_{\min} = 16.67$ms, respectively. The lower right ends of the lines correspond to $p_{\text{break}} = 0.93$ and $T_{\min} = 500$ms, respectively. The policies are identical when $p_{\text{break}} = 0.93$ and $T_{\min} = 16.67$ms.

Figure 10 compares the simulated performance of the three heuristic methods and the probabilistic method using the two fixation segmentation algorithms. For the heuristic models, we set $\overline{T} = 500ms$ and $q = 5$. We varied $\underline{T}$ from 16.67ms to 500ms in steps of 16.67ms. For the probabilistic method, we used Policy I and Policy II. Both the heuristic and the probabilistic methods achieve better tradeoffs between accuracy and speed than the uniform dwell time policy. However, the probabilistic methods achieve the best tradeoff, with the one with I-HMM2-based fixation segmentation outperforming the one with I-HMM-based fixation segmentation.

## Experimental Results: Objective Measures

Figure 8 compares the experimentally measured response times and error rates from Experiments II and III with the simulated results. Table 6 and Table 7 give the corresponding numerical data. In Figure 8, data from Experiment II is shown by the cross-hairs attached to the two circles on the dash-dotted red line (100ms uniform dwell time on the upper left and 300ms uniform dwell time on the lower right) and the cross-hairs attached to the circle with the second lowest error rate (Probabilistic with parameters [600ms, 100ms, 100ms, 1]). Data from Experiment III is shown by cross-hairs attached to the remaining circles. Moving from upper left to lower right along the pink solid and blue dashed lines, we show data from the five probabilistic policies listed in the section "Experimental Procedures." These correspond to Policy I with $p_{\text{break}}$=0.005, Policy I with $p_{\text{break}}$=0.3, Policy I with $p_{\text{break}}$=0.93 or equivalently Policy II with $T_{\text{min}}$=16.67ms, Policy II with $T_{\text{min}}$=100ms and Policy II with $T_{\text{min}}$=316.7ms.

The experimental results are quite similar to the corresponding simulated results. The simulated results usually lie within the 95% confidence intervals of the experimental results. The differences are partly due to differences between experimental conditions, subjects and tasks, as we mixed data from all three experiments in generating the simulation results. Slightly better matching between simulated and experimental results is obtained if we use only data from the same experiment in the simulations (data not shown).

Consistent with our simulated results, our experimental results show that Policies I and II significantly outperform the uniform dwell time policies. For instance, the variable dwell time Policy I with $p_{\text{break}}$=0.93 (equivalently Policy II with $T_{\text{min}}$=16.67ms) achieves a 50.7% lower error rate (16.16% vs 32.77%), while maintaining a similar average response time (198.82ms vs 168.77ms) when compared to the 100ms uniform policy. It achieves a similar error rate (16.16% vs 21.19%), but reduces the average response time by 60.9% (198.82ms vs 508.47ms) when compared to the 300ms uniform policy.

Figure 11 compares the response times and error rates for the uniform, heuristic and probabilistic policies, obtained both through simulation and by experiment (Experiment IV). Table 8 gives the corresponding numerical data. The parameters for the policies were chosen so that response times predicted by simulation were all approximately the same (~300ms).

Experimentally measured response times for the different policies were all similar to each other, but lower (~260ms) than predicted in simulation (~300ms). A repeated measures ANOVA did not indicate a statistically significant effect of dwell time policies on the response time ($F(5, 55) = 0.79$, $p = 0.56$).

All variable dwell time policies had lower error rates than the uniform dwell time policy. The ranking of the policies in terms of error rate was preserved when going from simulation to experiment, except for the temporal and distance-based heuristic methods, which had quite similar results. The heuristic methods yielded relatively small decreases from uniform dwell time policy in error rate (3.96%, 9.01% and 10.00% for the distance-based, temporal and duration-based methods respectively). The probabilistic method yielded the largest reduction in error rate (14.89% with I-HMM-based fixation segmentation and 18.16% with I-HMM2-based fixation segmentation). A repeated measures ANOVA indicated a statistically significant effect of dwell time policies on error rate ($F(5, 55) = 4.24$, $p < .01$). Post-hoc test indicated the differences in error rate when comparing the probabilistic method with I-HMM2 with the uniform dwell time policy ($p < .01$) and heuristic distance-based policy ($p = .01$) were significant.

**Experimental Results: Subjective Measures**

In Experiment II, we compared subjects' subjective perceptions about the accuracy and response speeds of three uniform dwell time policies and a probabilistic variable dwell time policy.

Subjects were asked to assign a score to each policy according to the accuracy and response speed. Scores from each individual subject were mean-centered by subtracting the mean scores given by the subject. Figure 12 plots the mean-centered scores, where the directions of the axes were reversed so that the plot is comparable to Figure 8, i.e. lower accuracy/higher error rate in the upper part of the graph and slower speed/longer response time on the right hand side. Similar to the objective measurements, increases in the dwell time lead to subjective improvements in accuracy but decreases in perceived speed. In addition, the point for the variable dwell time policy lies to the lower left of the line connecting the uniform dwell time policies, indicating that subjects felt that variable dwell time enables a better tradeoff between speed and accuracy.

Statistical tests indicated that the observed differences in both perceived speed and accuracy were significant. A repeated measures ANOVA indicated a statistically significant effect of dwell time policies on the perceived response speed ($F(3, 33) = 27.91$, $p < .0001$). Post-hoc tests indicated significant differences in perceived speed when comparing the variable dwell time policy with the 100ms ($p < .01$) and 500ms ($p < .01$) policies, but not with the 300ms policy ($p = .42$). Similarly, a repeated measures ANOVA indicated a statistically significant effect of dwell time policies on the perceived accuracy ($F(3, 33) = 12.76$, $p < .0001$). Post-hoc test indicated significant differences in perceived accuracy when comparing the variable dwell time policy with the 100ms ($p < .001$) and 300ms ($p = .04$) policies, but not with the 500ms policy ($p = .95$).

Subjects were also asked to rank the policies according to their overall preference. Figure 13 presents histograms of the ranks assigned to the different dwell time policies. Consistent with previous results, subjects tended to assign higher ranks to the variable dwell time policy. The means and standard deviations of the rankings were $3.33 \pm 0.98$ (100ms), $2.58 \pm 1.08$ (300ms), $2.58 \pm 1.00$ (500ms) and $1.50 \pm 0.67$ (variable). Under two-tailed Wilcoxon signed-rank tests, the ranking of variable dwell time policy was statistically significantly different from the ranking of the three uniform dwell time policies (100ms, $Z = 2.87$, $p < .01$; 300ms, $Z = 2.17$, $p = .03$; 500ms, $Z = 2.18$, $p = .03$).

In Experiment IV, we asked subjects to compare six policies: one uniform, three heuristic and two probabilistic policies, all with similar response times predicted by simulation. Figure 14 shows the mean-centered scores for perceived accuracy and perceived response speed. As expected, the perceived response speeds of the different policies are very similar. A repeated measures ANOVA did not indicate a significant effect of dwell time policies on perceived response speed ($F(5, 55) = 1.94$, $p =.10$). On the other hand, the different policies had statistically significant differences in perceived accuracy according to a repeated measures ANOVA ($F(5, 55) = 4.85$, $p < .001$). All variable dwell time policies received higher scores for accuracy than the uniform policy, with the probabilistic method with I-HMM2-based fixation segmentation receiving the highest average score. Post-hoc test indicated significant differences in perceived accuracy when comparing the probabilistic policy with I-HMM2-based segmentation with the uniform dwell time policy ($p < .01$) and the heuristic distance-based policy ($p = .02$).

Figure 15 presents histograms of the ranks assigned to the different policies. All variable policies received higher average ranking than the uniform dwell time policy. The probabilistic with I-HMM2-based fixation segmentation receives the highest average ranking. The mean and standard deviations of the ranks were $4.67 \pm 1.23$ (uniform), $4.50 \pm 1.62$ (heuristic distance-based), $3.92 \pm 1.56$ (heuristic temporal), $3.42 \pm 1.73$ (heuristic duration-based), $2.50 \pm 1.57$ (probabilistic with I-HMM) and $2.00 \pm 0.85$ (probabilistic with I-HMM2). Under two-tailed Wilcoxon signed-rank tests, the ranking of probabilistic policy with I-HMM2 was statistically significantly different from the ranking of the uniform dwell time policy ($Z$=3.08, $p$<.01), the heuristic distance-based method ($Z$=2.88, $p$<.01) and the heuristic temporal method ($Z = 2.61$, $p < .01$).

## DISCUSSION

Past work in gaze-based selection has typically used a single dwell time applied uniformly across all objects to avoid the “Midas touch” problem. The choice of this dwell time is a tradeoff between response speed and error rate. Shorter dwell times lead to shorter responses, but more frequent errors.

To achieve a better tradeoff, we propose to assign different dwell times to different objects. In particular, we studied several methods to assign variable dwell times to different hyperlinks in a gaze-based web browser, which used a two-step selection policy. Intuitively, hyperlinks which are more likely to be selected are assigned shorter dwell times.

We have introduced a new probabilistic method for inferring the likelihood that the user wishes to select each hyperlink on the page. The probabilistic method is based upon a statistical generative model of natural gaze behavior during web browsing. This model enables us to incorporate both spatial and temporal information about the gaze trajectories in a principled way. It contains two stages. In the first stage, the past gaze trajectory is segmented into a sequence of fixations. In the second stage, the model uses this sequence to infer the probability that each hyperlink on the page is the one the user would like to select. Each stage is based upon a hidden Markov model whose model parameters can be identified using training data collected from users as they surf the web using a gaze-based web browser with a uniform dwell time.

Based on the inferred likelihoods, we proposed a four-parameter class of policies where the dwell time of a link decreased in a piecewise linear manner with the inferred probability that the link is the user’s desired selection. We performed extensive simulations to identify the policy parameters that resulted in the best tradeoff between accuracy and response speed, and identified two single parameter policies which could achieve the best tradeoff. Our simulation results suggested that these probabilistic variable dwell time policies would achieve better tradeoffs than using a uniform dwell time policy.

Our experimental results evaluating this probabilistic method both objectively (Figure 8, Table 6 and Table 7) and subjectively (Figure 12 and Figure 13) confirmed this expectation. Our probabilistic variable dwell time policy achieves a similar response time as a 100ms uniform dwell time policy, but reduces errors by 50%. It achieves a similar accuracy as a 300ms uniform dwell time policy, but reduces response time by 60%.

We also compared our proposed method with the other variable dwell time policies based on three simple heuristics (decreasing dwell time with distance from the last fixation, decreasing dwell time according to the order of fixation, and decreasing dwell time with the duration of fixation) both objectively (Figure 11, Table 8) and subjectively (Figure 14 and Figure 15). All variable dwell time policies had similar measured and perceived response times as a uniform 183.33ms dwell time policy, but lower measured and perceived error rates, as well as higher overall average rank.

Our proposed probabilistic model achieves the lowest measured and perceived error rates and the highest subjective ranking among all policies considered. However, our results show that even simple heuristic methods for varying dwell time can lead to improved user experience. Heuristic methods are appealing, because they are simple, intuitive and easily understandable. Because our examination of heuristic methods is by no means exhaustive, we cannot rule out the possibility that another heuristic method might result in better performance than our proposed probabilistic method. Nonetheless, there are a number of additional reasons why we believe the probabilistic approach described here is a more promising line of investigation.

First, the probabilistic model integrates spatial information in a more mathematically principled way than the heuristic methods studied. The use of Gaussian densities for the spatial distribution of gaze in the probabilistic model enables it to account both for noise in the eye tracking and for the possibility that the hyperlink of interest may not be directly fixated. Experiment I showed that the top-5 accuracy of the probabilistic model is 8.14% greater than the top-5 accuracy of the temporal and duration-based heuristic models, which considered only the last five directly fixated hyperlinks (Figure 7(a)). This implies the probabilistic model can correctly infer the desired hyperlink, even if it is not directly fixated upon. This is especially advantageous for web browsing, since the hyperlinks can be close, and their layout can vary from page to page.

Second, the model also benefits from the principled incorporation of temporal information. The past history of gaze is taken into account by the sequence modelling provided by the hidden Markov models, which also enables us to model different types of gaze behavior. Fixation duration is accounted for by the emission densities. Experiment I also shows the advantages of including this temporal information. The top-*k* accuracy of the probabilistic model consistently exceeds that of the distance-based model, which took into account spatial information, but considered only the last fixation point, by at least 9.30% (Figure 7(a)).

We acknowledge that heuristic methods might be extended to incorporate more cues. However, we argue that the additional complexity incurred would negate much of the simplicity that makes them so attractive.

Third, the process of finding many of the model parameters can be automated using machine learning, based on data collected from users as they browse the web using a gaze based browser using a uniform dwell time policy. This reduces bias that might be introduced by prior knowledge of the experimenter. We feel that the proposed models are accurate models of gaze behavior, as the learned distributions are consistent with experimental findings from the psychophysical literature. For example, it has been found that long fixations are usually related to user's interest (Orquin & Loose, 2013). This is consistent with our finding in Figure 5 the learned distribution of fixation durations is biased towards longer values when the user is reading the hyperlink, than when the user is reading away from the hyperlink, and with our finding that the distribution is biased towards intermediate values when the user is reading near the hyperlink.

Fourth, in the future the models can be customized to individual users by tuning the model parameters based on that user's gaze data, leading to potential further improvements. However, we emphasize that individualized tuning is not necessary to achieve the performance gains here. Our experimental results with leave-one-subject-out cross validation in Figure 7 show that even when trained on data from other subjects, the probabilistic model can more accurately identify the hyperlink the user wishes to select than any of the three heuristic methods. We also note that most of the subjects in Experiment IV comparing the different gaze models were not seen in any of the prior experiments, so their data was not used in training the models tested.

Moving forward, we plan to extend this method by taking more information into account in the intent inference model, e.g., the information of the users' previous command (Salvucci & Anderson, 2000) and information about the current task. Because we have formulated our model probabilistically, we anticipate that with the proper formulation, we will be able to incorporate this information in a mathematically principled and rigorous manner.

The basic framework we have used here, two-step dwell-based selection, is commonly used in current state-of-the-art systems, such as Windows Control and GazeTheWeb. Thus, our algorithm can be adopted with minimal change to many existing interfaces. Because the basic paradigm for selection is unchanged, our proposed method does not increase cognitive load on the users, and will require little if any retraining.

We also plan to evaluate this method in more general selection tasks or other dwell time based control strategies, beyond the two-step selection policy for hyperlink selection studied here. We anticipate that adaptive adjustment of dwell time based on past gaze history will improve the performance in other contexts as well.

The technique of adaptive dwell time is complementary to other approaches for improving gaze-based interfaces, such as fisheye lenses (Ashmore, Duchowski, & Shoemaker, 2005), automatic magnification (2017; Menges, Kumar, Sengupta, & Staab, 2016), and real-time visual feedback (C. Kumar, Menges, & Staab, 2016). Thus, we expect that the combination of our algorithm and these techniques can yield more adaptive and stable eye-based systems with even better performance.

Author Note

**Zhaokang Chen** received the B.S. degree from Fudan University, China, in 2014. He is currently pursuing Ph.D. degree in the Department of Electronic and Computer Engineering at the Hong Kong University of Science and Technology. His research interests include human computer interaction, brain computer interface, eye tracking, and machine learning.

**Bertram E. Shi** is a professor in the Departments of Electronic and Computer Engineering and Chemical and Biological Engineering at the Hong Kong University of Science and Technology. His research interests are in bio-inspired visual processing and robotics, neuromorphic engineering, computational neuroscience, machine learning, and developmental robotics.

# Tables

*Table 1*

$\mu_e(l_{t-2}l_{t-1}g_{t-2}g_{t-1})$

| $l_{t-2}$ \ $l_{t-1}$ | *f* | *s* | *o* |
|---|---|---|---|
| *f* | $(g_{t-2}+g_{t-1})/2$ | $g_{t-1}$ | $g_{t-2}$ |
| *s* | $(g_{t-2}+g_{t-1})/2$ | $g_{t-1}$ | |
| *o* | $g_{t-1}$ | | |

*Table 2*

$\Sigma_e(l_{t-2}l_{t-1}l_t)$ when $l_t = f$

| $l_{t-2}$ \ $l_{t-1}$ | $f$ | $s$ | $o$ |
|---|---|---|---|
| $f$ | $1.5\Sigma_f$ | $2\Sigma_f$ | $2\Sigma_f$ |
| $s$ | $1.5\Sigma_f$ | $2\Sigma_f$ | |
| $o$ | $2\Sigma_f$ | | |

*Table 3*

Participation Information of 34 Subjects

| Experiment \ # of subjects | 8 | 4 | 4 | 9 | 2 | 2 | 2 | 1 | 1 | 1 | Total |
|---|---|---|---|---|---|---|---|---|---|---|---|
| I | * | | | | * | | * | * | * | | 14 |
| II | | * | | | * | * | * | * | | * | 12 |
| III | | | * | | | * | * | | * | * | 10 |
| IV | | | | * | | | | * | * | * | 12 |

*Table 4*

Averages and Standard Deviations of the FHMM Transition Matrix Probabilities

| $B_{t-1}$ \ $B_t$ | 1 | 2 | 3 | TS |
|---|---|---|---|---|
| 1 | 0.57 ± 0.03 | 0.00 ± 0.00 | 0.08 ± 0.01 | 0.35 ± 0.03 |
| 2 | 0.34 ± 0.03 | 0.55 ± 0.02 | 0.03 ± 0.01 | 0.08 ± 0.01 |
| 3 | 0.05 ± 0.02 | 0.16 ± 0.03 | 0.59 ± 0.02 | 0.20 ± 0.01 |

*Table 5*

Averages and Standard Deviations of the FHMM Emission Distribution Parameters

| $\beta_{x1}$ | $\beta_{y1}$ | $\beta_{x2}$ | $\beta_{y2}$ | $\mu_{d1}$ | $\mu_{d2}$ | $\mu_{d3}$ | $\pi_1$ | $\pi_2$ |
|---|---|---|---|---|---|---|---|---|
| 0.17 ± 0.02 | 0.48 ± 0.15 | 0.00* ± 0.00* | 0.00* ± 0.00* | 2.90 ± 0.02 | 2.64 ± 0.02 | 2.54 ± 0.02 | 0.08 ± 0.05 | 0.66 ± 0.05 |
| $\sigma_{x1}$ (px) | $\sigma_{y1}$ (px) | $\sigma_{x2}$ (px) | $\sigma_{y2}$ (px) | $\sigma_{d1}$ | $\sigma_{d2}$ | $\sigma_{d3}$ | $\pi_3$ | |
| 39.38 ± 2.42 | 14.07 ± 1.51 | 126.43 ± 12.92 | 38.41 ± 2.14 | 0.62 ± 0.02 | 0.45 ± 0.01 | 0.47 ± 0.01 | 0.26 ± 0.03 | |

*Note*: * These values were smaller than 5e-3.

*Table 6*

Experimental Results of Experiment II and Simulated Results

| | Response Time (ms) | | Error Rate | |
|---|---|---|---|---|
| | Simulated | Experimental | Simulated | Experimental |
| 100ms uniform policy | 178.27 ± 8.66 | 168.77 ± 7.54 | 32.03% ± 5.24% | 32.77% ± 8.47% |
| 300ms uniform policy | 473.31 ± 17.57 | 508.47 ± 38.59 | 16.67% ± 4.18% | 21.19% ± 7.40% |
| [600ms, 100ms, 100ms, 1] | 388.29 ± 29.82 | 372.08 ± 52.77 | 12.42% ± 3.70% | 10.00% ± 5.39% |
| 500ms uniform policy | 733.55 ± 22.78 | 730.83 ± 38.35 | 11.11% ± 3.53% | 13.33% ± 6.11% |

*Note*: Each entry is presented as Mean ± 95% CI (confidence interval)

*Table 7*

Experimental Results of Experiment III and Simulated Results

| | Response Time (ms) | | Error Rate | |
|---|---|---|---|---|
| | Simulated | Experimental | Simulated | Experimental |
| [500ms,16.67ms, 50ms , 0.0005] | 90.34 ± 13.71 | 72.33 ± 10.68 | 28.43% ± 5.06% | 30.00% ± 9.03% |
| [500ms, 16.67ms, 50ms , 0.3] | 147.93 ± 20.89 | 109.26 ± 21.99 | 18.63% ± 4.37% | 23.23% ± 8.36% |
| [500ms, 16.67ms, 16.67ms, 1] | 274.24 ± 27.94 | 198.82 ± 38.35 | 13.40% ± 3.82% | 16.16% ± 7.29% |
| [500ms, 100ms, 100ms, 1] | 385.51 ± 27.63 | 304.33 ± 34.34 | 13.07% ± 3.78% | 11.00% ± 6.16% |
| [500ms, 316.7ms, 316.7ms, 1] | 594.06 ± 24.31 | 524.00 ± 40.63 | 9.48% ± 3.29% | 7.00% ± 5.03% |
| 500ms uniform policy | 733.55 ± 22.78 | 728.83 ± 38.69 | 11.11% ± 3.53% | 13.00% ± 6.62% |

*Note*: Each entry is presented as Mean ± 95% CI (confidence interval)

*Table 8*

Experimental Results of Experiment IV and Simulated Results

| | Response Time (ms) | | Error rate | |
|---|---|---|---|---|
| | Simulated | Experimental | Simulated | Experimental |
| Uniform dwell time | 305.23 ± 14.31 | 282.08 ± 16.23 | 26.47% ± 4.95% | 28.33% ± 8.10% |
| Heuristic distance-based | 301.63 ± 25.99 | 240.06 ± 31.22 | 21.24% ± 4.59% | 24.37% ± 7.75% |
| Heuristic temporal | 297.44 ± 22.37 | 249.58 ± 28.15 | 19.93% ± 4.48% | 19.33% ± 7.12% |
| Heuristic duration-based | 296.84 ± 21.77 | 254.31 ± 32.25 | 20.26% ± 4.51% | 18.33% ± 6.95% |
| Probabilistic with IHMM | 308.28 ± 31.17 | 279.83 ± 42.42 | 15.36% ± 4.05% | 13.45% ± 6.16% |
| Probabilistic with IHMM2 | 302.23 ± 28.28 | 255.93 ± 39.28 | 13.73% ± 3.86% | 10.17% ± 5.48% |

*Note*: Each entry is presented as Mean ± 95% CI (confidence interval)

# Figures

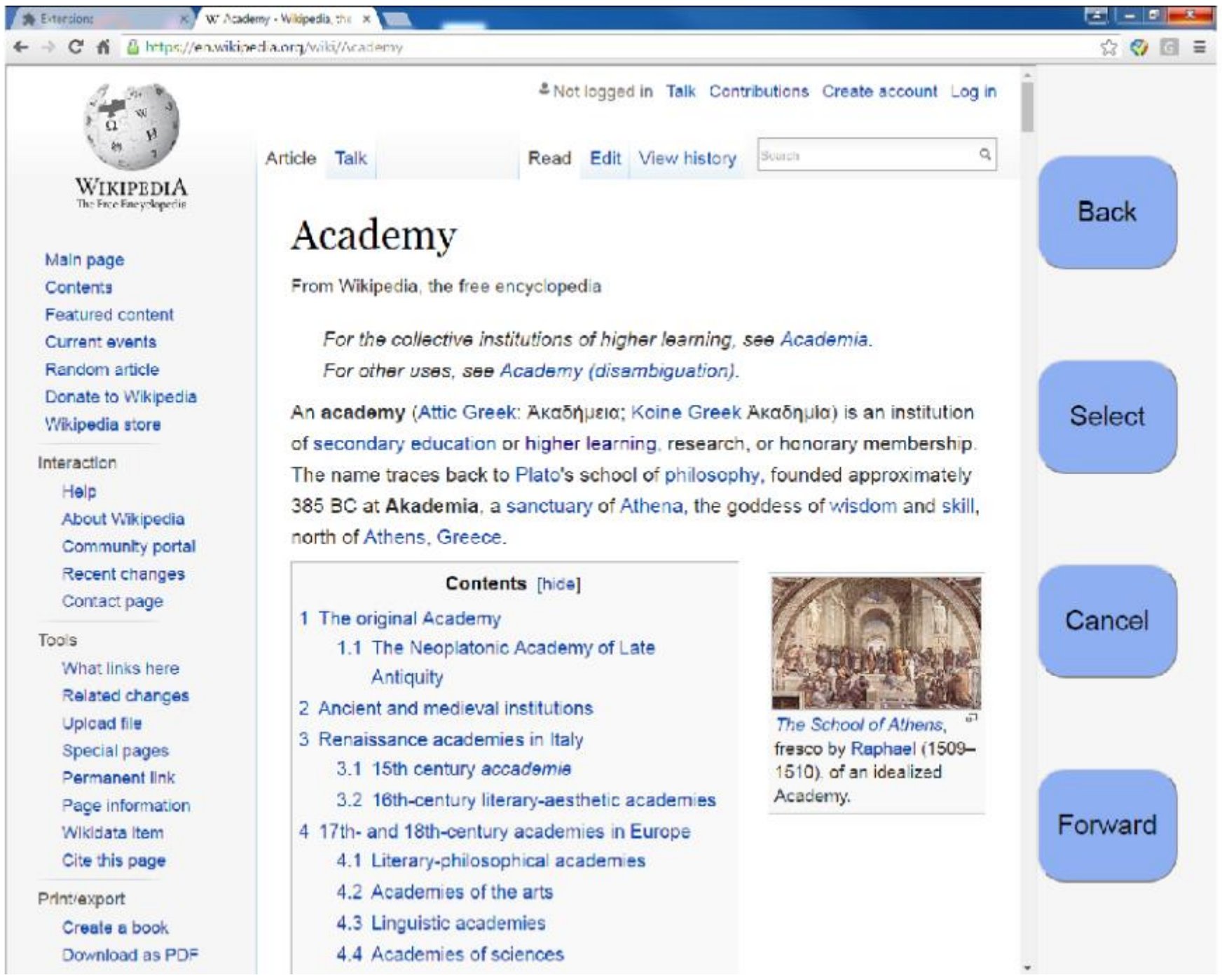

*Figure 1*. The graphical user interface used in this study. There are four command buttons: back, select, cancel and forward. To select a hyperlink, a two-step selection policy is used: the users first maintain their gaze inside the "Select" button to activate it, and then maintain their gaze on the hyperlink they wish to select.

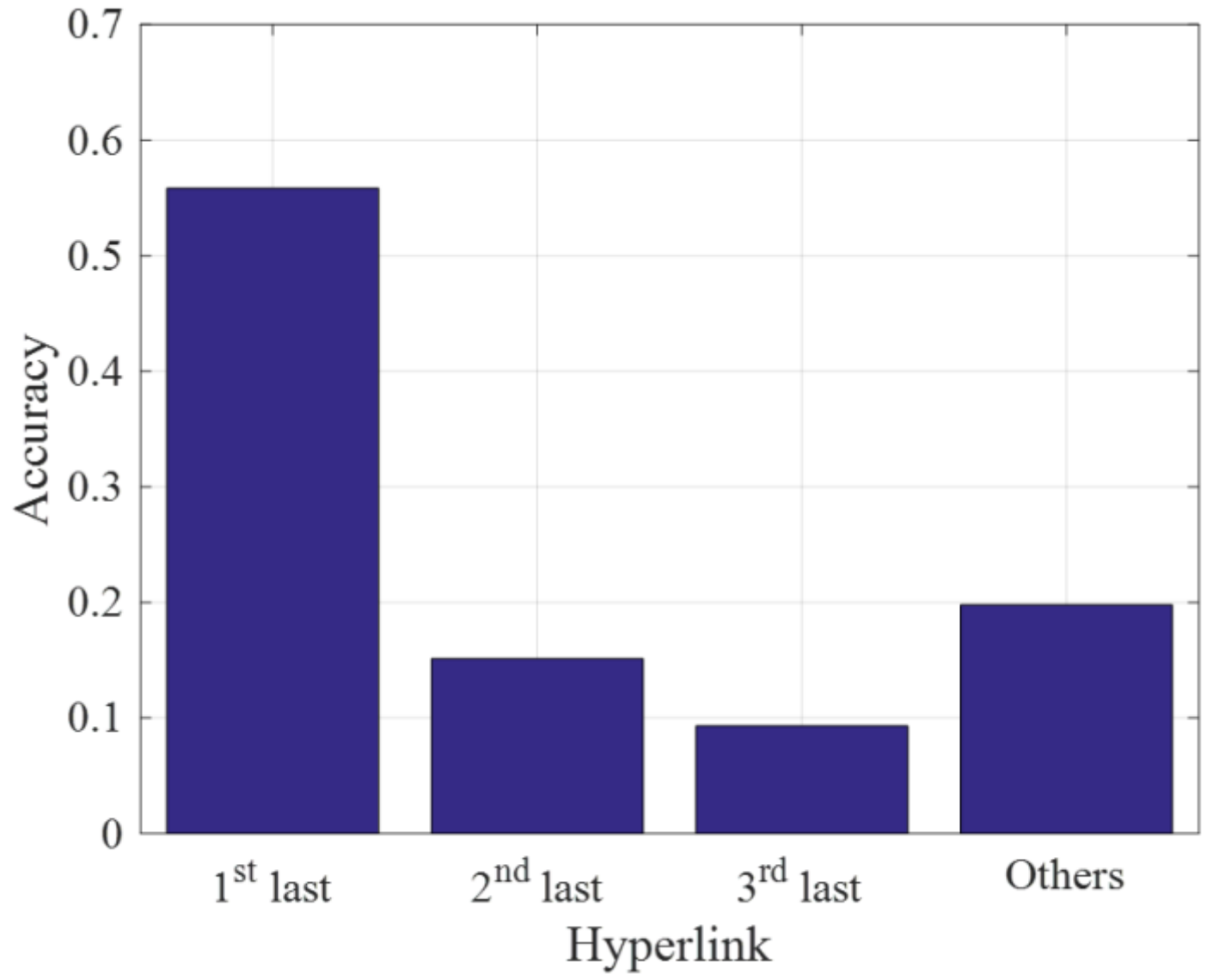

*Figure 2*. Percentage of the time the selected hyperlink was among the last three fixations before the "Select" button was activated.

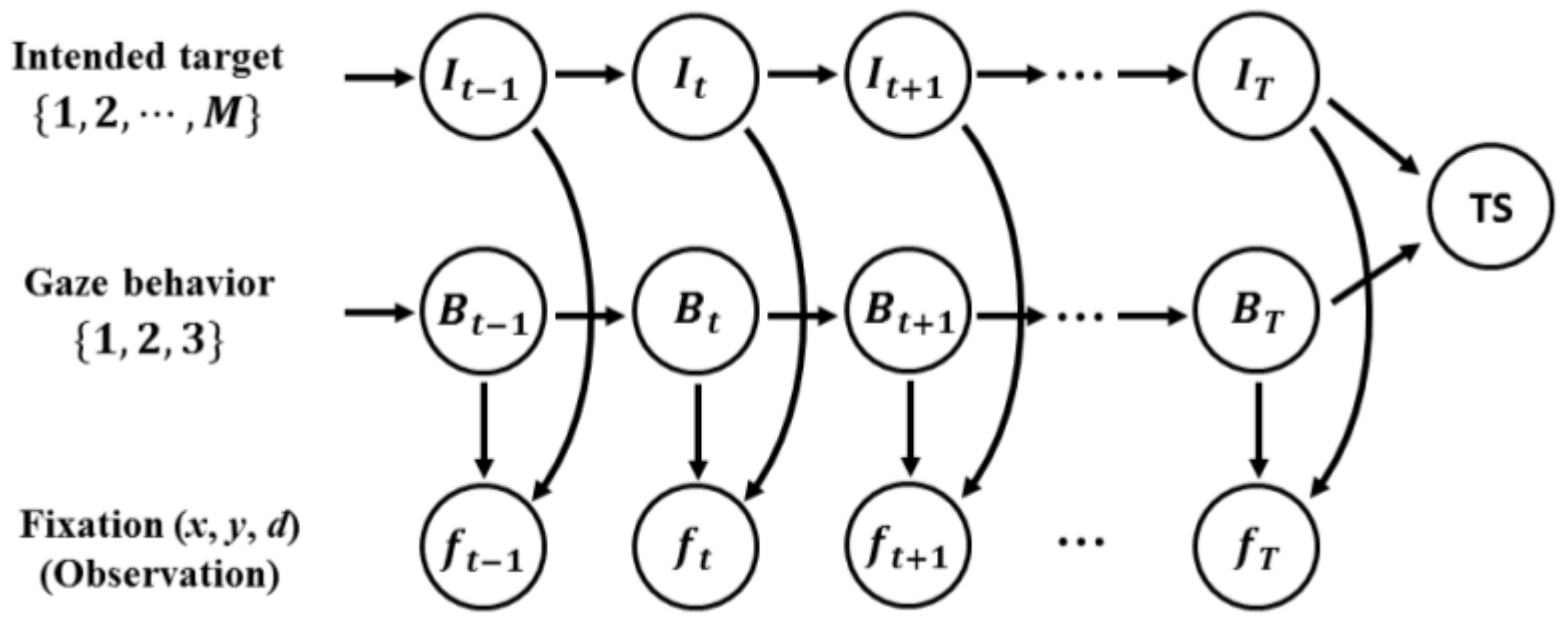

*Figure 3*. Bayesian network representing the factorial hidden Markov model used in the probabilistic eye gaze model. Observations $f_t$ depend upon two hidden states, the intended target $I_t$ and the gaze behavior, $B_t$. The terminal state (TS) indicates the activation of the “Select” button.

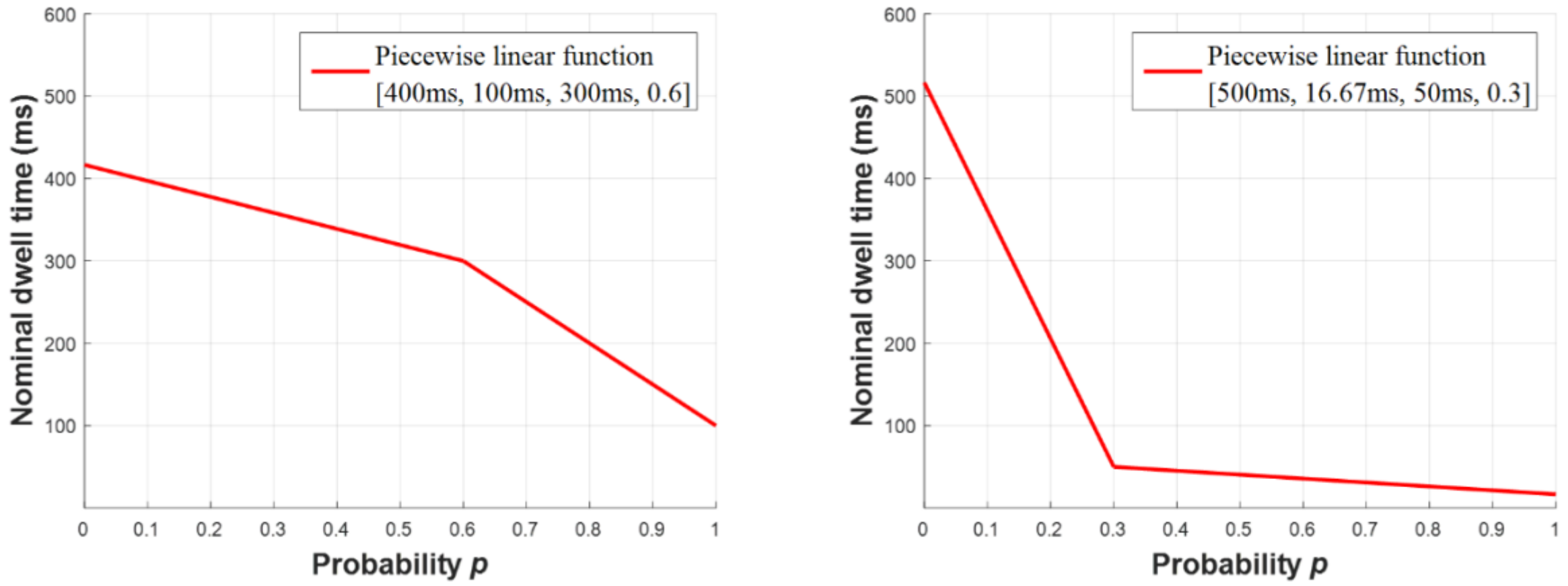

(a) (b)

*Figure 4*. Examples of the piecewise linear dwell time assignment policies with parameters [$T_{\text{max}}$, $T_{\text{min}}$, $T_{\text{break}}$, $p_{\text{break}}$] given by (a) [400ms, 100ms, 300ms, 0.6] and (b) [500ms, 16.67ms, 50ms, 0.3].

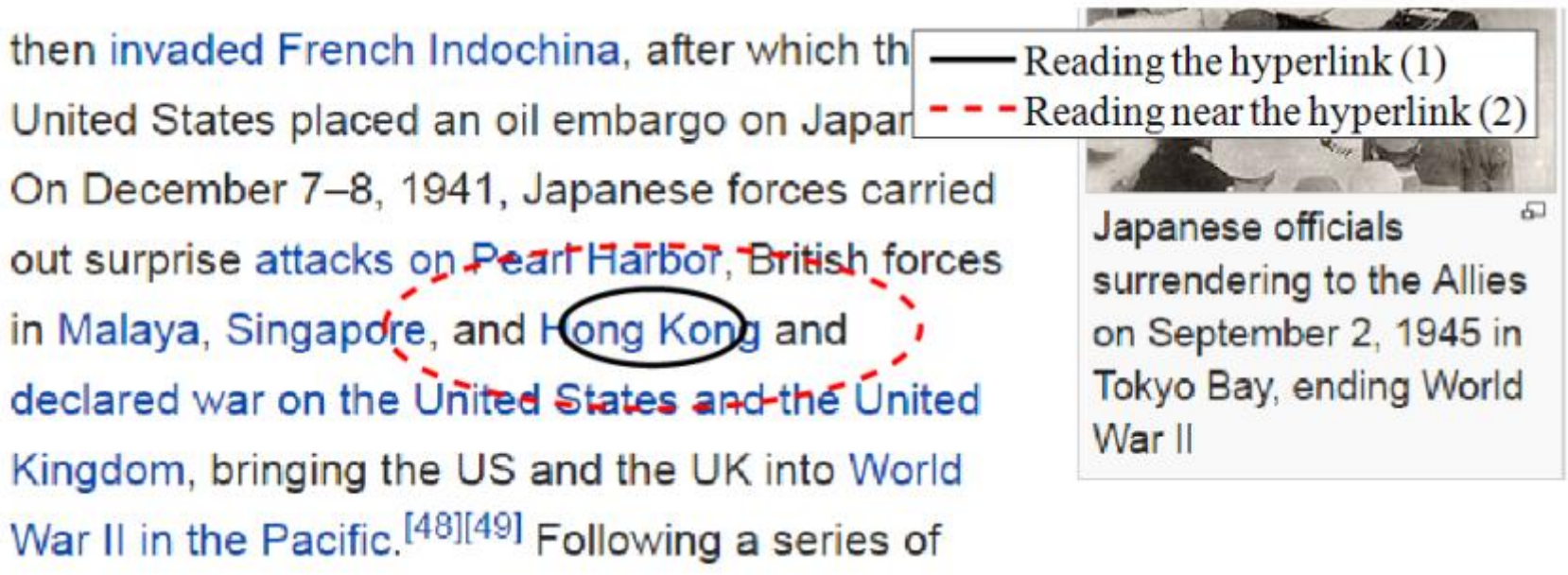

(a)

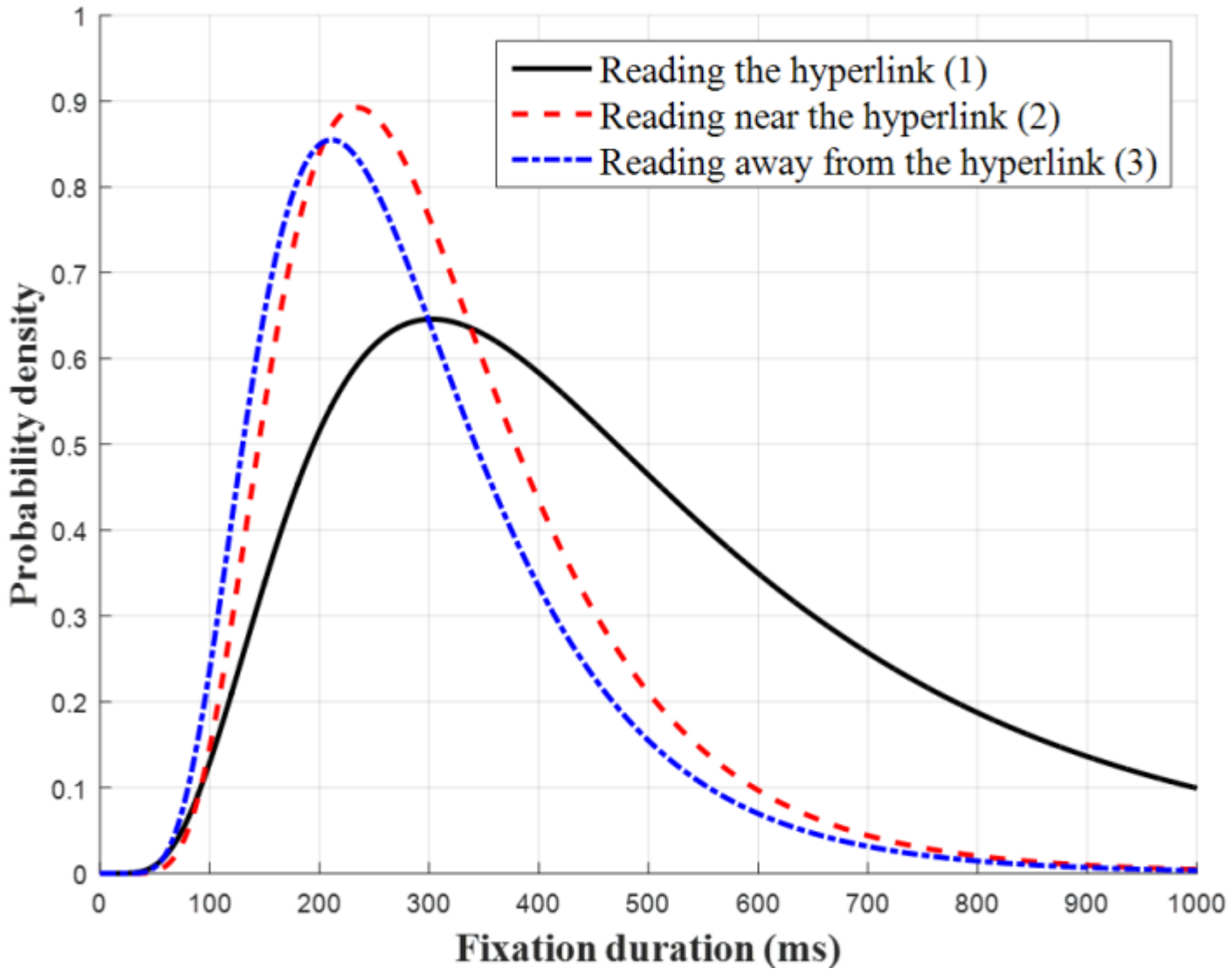

(b)

*Figure 5*. The emission densities from the FHMM scanpath model. (a) The two covariance ellipses of the Gaussian emission densities describing the fixation location assuming that the target hyperlink is “Hong Kong” under the conditions that the subject is looking at the hyperlink (1) and looking near the hyperlink (2). (b) The emission densities describing the fixation duration.

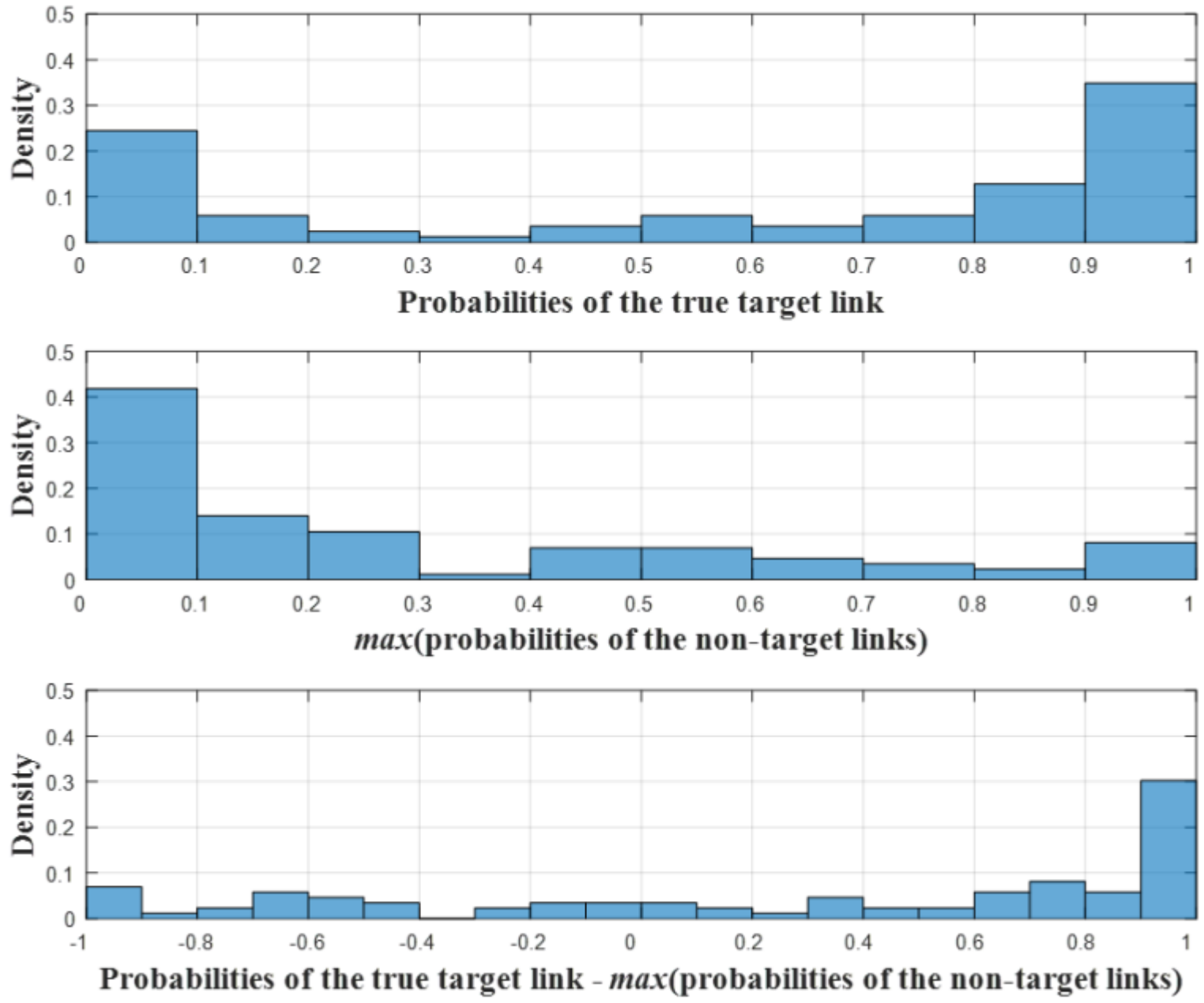

*Figure 6*. The histograms of the probabilities inferred from the FHMM scanpath model. The top plot shows the histogram of the inferred probability of the true target link. The middle plot shows the histogram of the highest probability among the non-target links. The bottom plot shows the histogram of the differences between them.

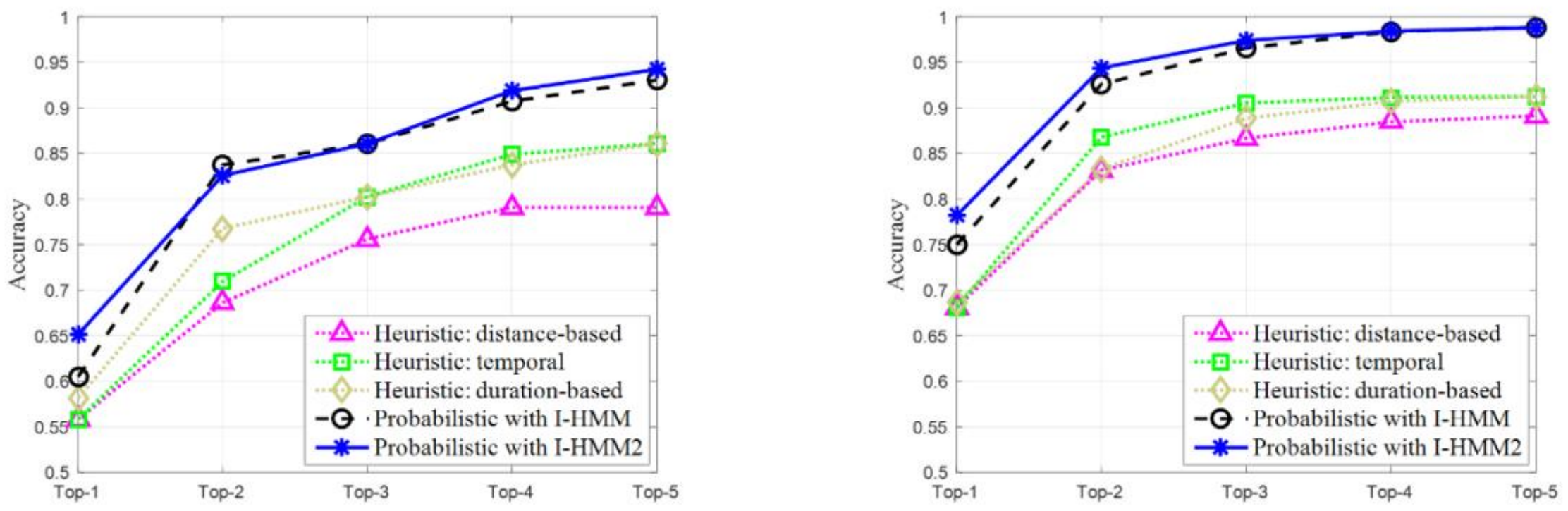

(a) Experiment I (b) Experiment II, III and IV

*Figure 7*. Accuracies of the rankings given by the heuristic and probabilistic methods. (a) Average accuracy computed over the fourteen folds using the data from Experiment I. (b) Accuracy computed over the data in Experiments II, III and IV using a model trained from all data in Experiment I.

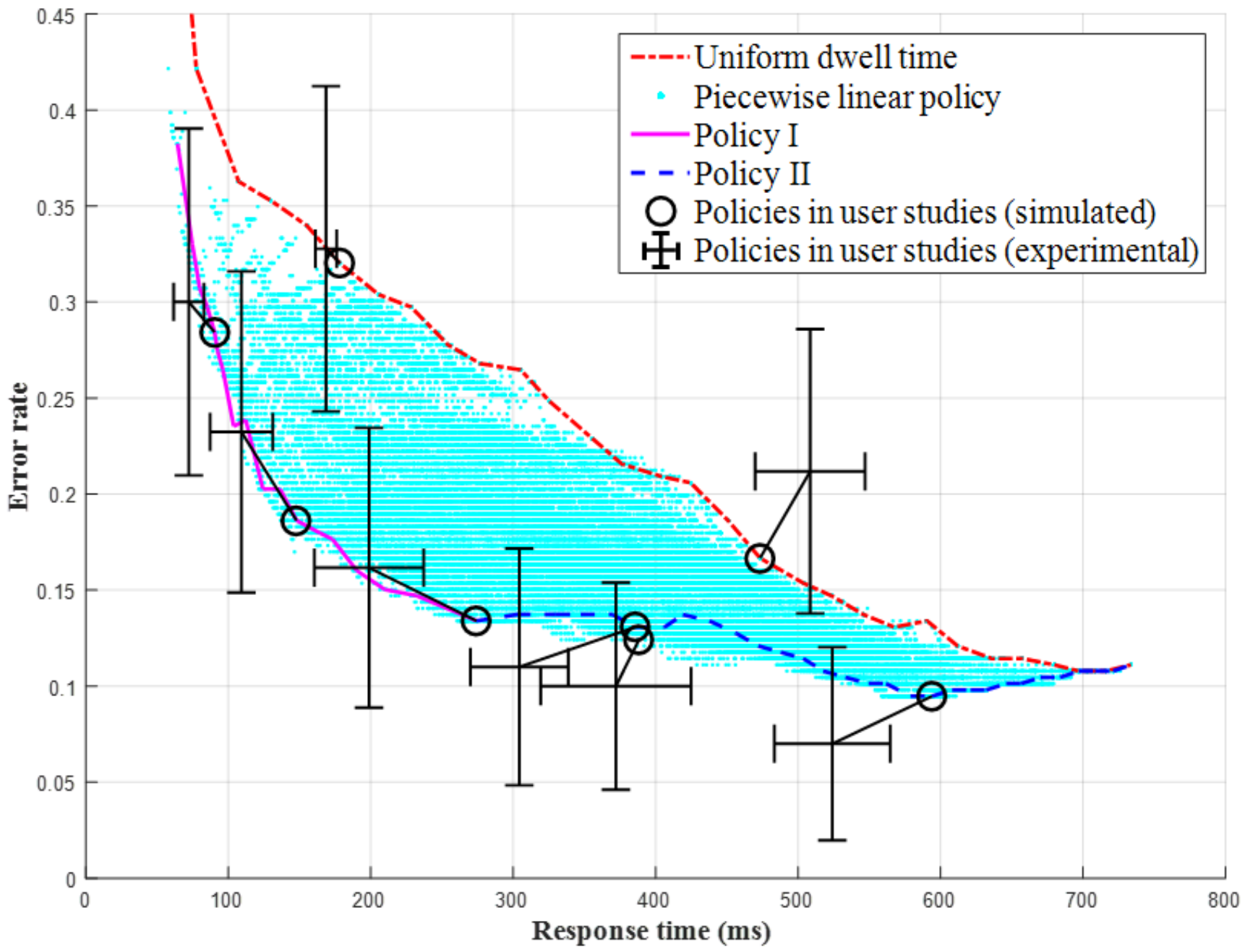

*Figure 8*. Simulated and experimental results from different dwell time assignment policies. Each cyan dot shows the simulated error rate and simulated response time for policies in the exhaustive search. The lines show the simulated performance of a class of policies: uniform dwell time (red dash-dotted), policy I (pink solid) and policy II (blue dashed). The center of each cross shows the experimental performance of a specific policy, which are connected to the simulated performance of the same policy (circle) by a line. The lengths of the cross-hairs indicate 95% confidence intervals.

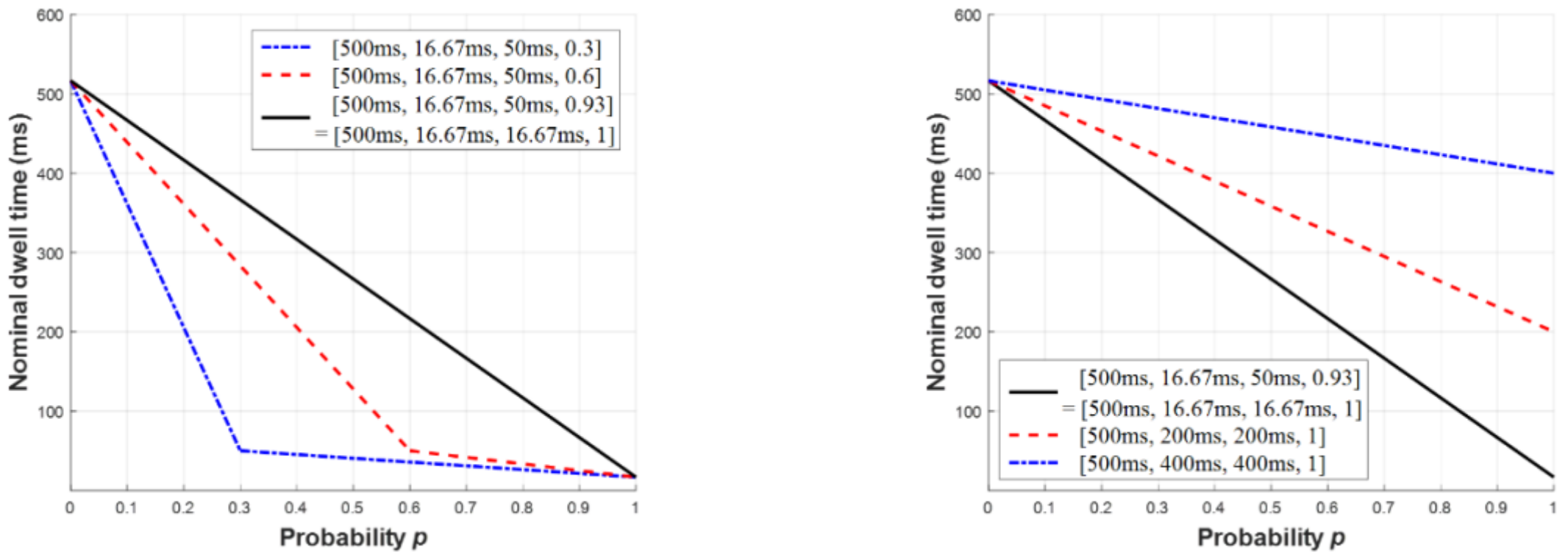

(a) (b)

*Figure 9*. Illustrations of the two single parameter policies that approximate the best tradeoff achieved by the piecewise linear policies. (a) Policy I, $T_{max}$ = 500ms, $T_{min}$ = 16.67ms, $T_{break}$ = 50ms, $p_{break} \in [0, 0.93]$; and (b) Policy II, $T_{max}$ = 500ms, $T_{min} = T_{break} \in$ [16.67ms, 500ms], $p_{break}$ = 1.

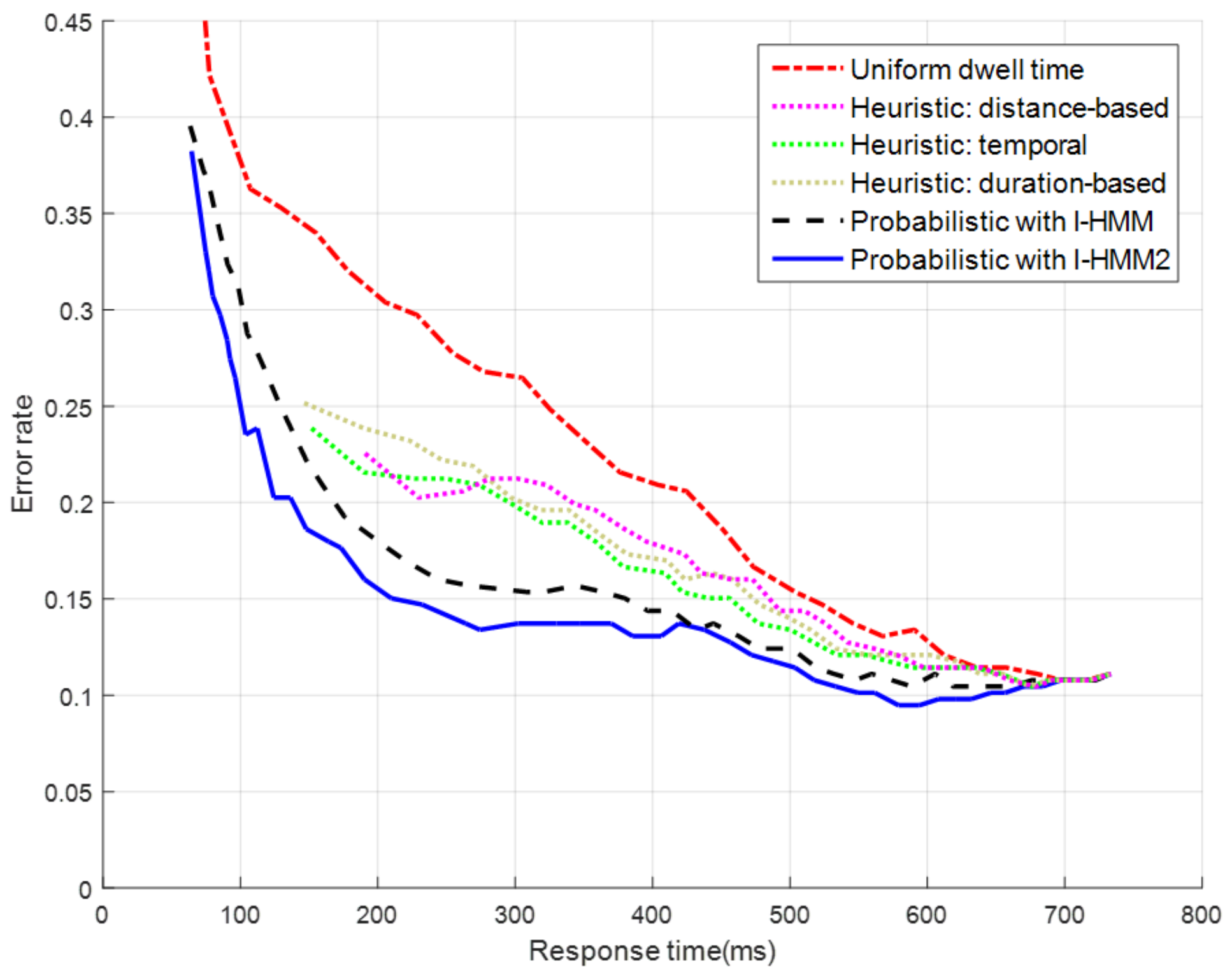

*Figure 10*. Simulated results from different target link inference models. Each line shows the simulated performance of an inference model: uniform dwell time (red dash-dotted), heuristic distance-based ranking (pink dotted), heuristic temporal ranking (green dotted), heuristic duration-based ranking (yellow dotted), probabilistic with I-HMM based fixation segmentation (black dashed), and probabilistic with I-HMM2-based fixation (blue solid).

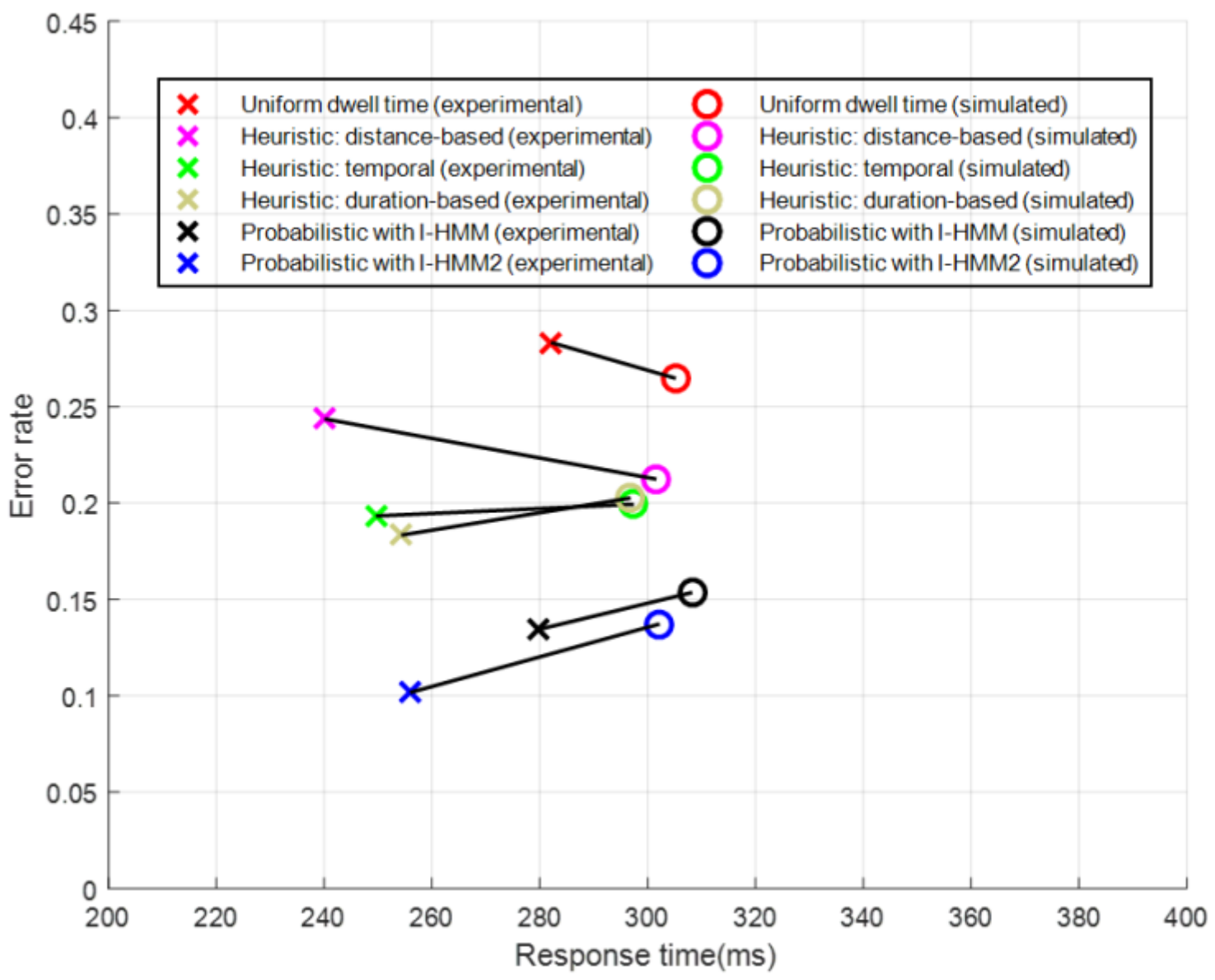

*Figure 11*. Simulation and experimental results from different target link inference models. Each marker shows the selection performance of one target link inference models: uniform dwell time (red), heuristic distance-based ranking (pink), heuristic temporal ranking (green), heuristic duration-based ranking (yellow), probabilistic with I-HMM-based fixation segmentation (black) and probabilistic with I-HMM2-based fixation segmentation (blue). Cross markers present experimental results. Circle markers present simulation results.

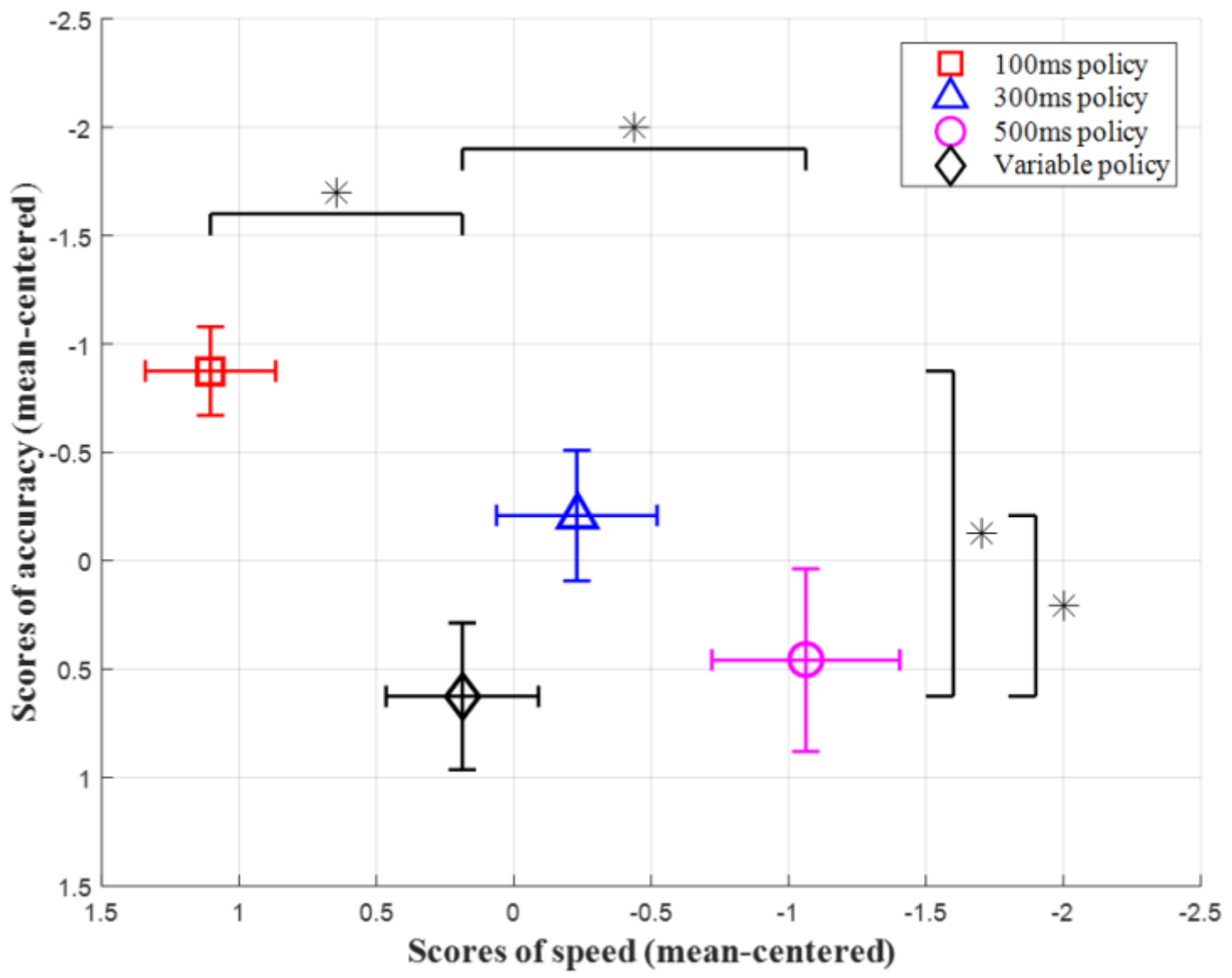

*Figure 12*. The scores of accuracy and speed collected by questionnaires in Experiment II. The scores of each individual subject are mean-centered by subtracting the mean scores of the subject. The error bars represent 95% confidence intervals. The square brackets with asterisks indicate differences with the variable dwell time policy that are statistically significant ($p < .05$).

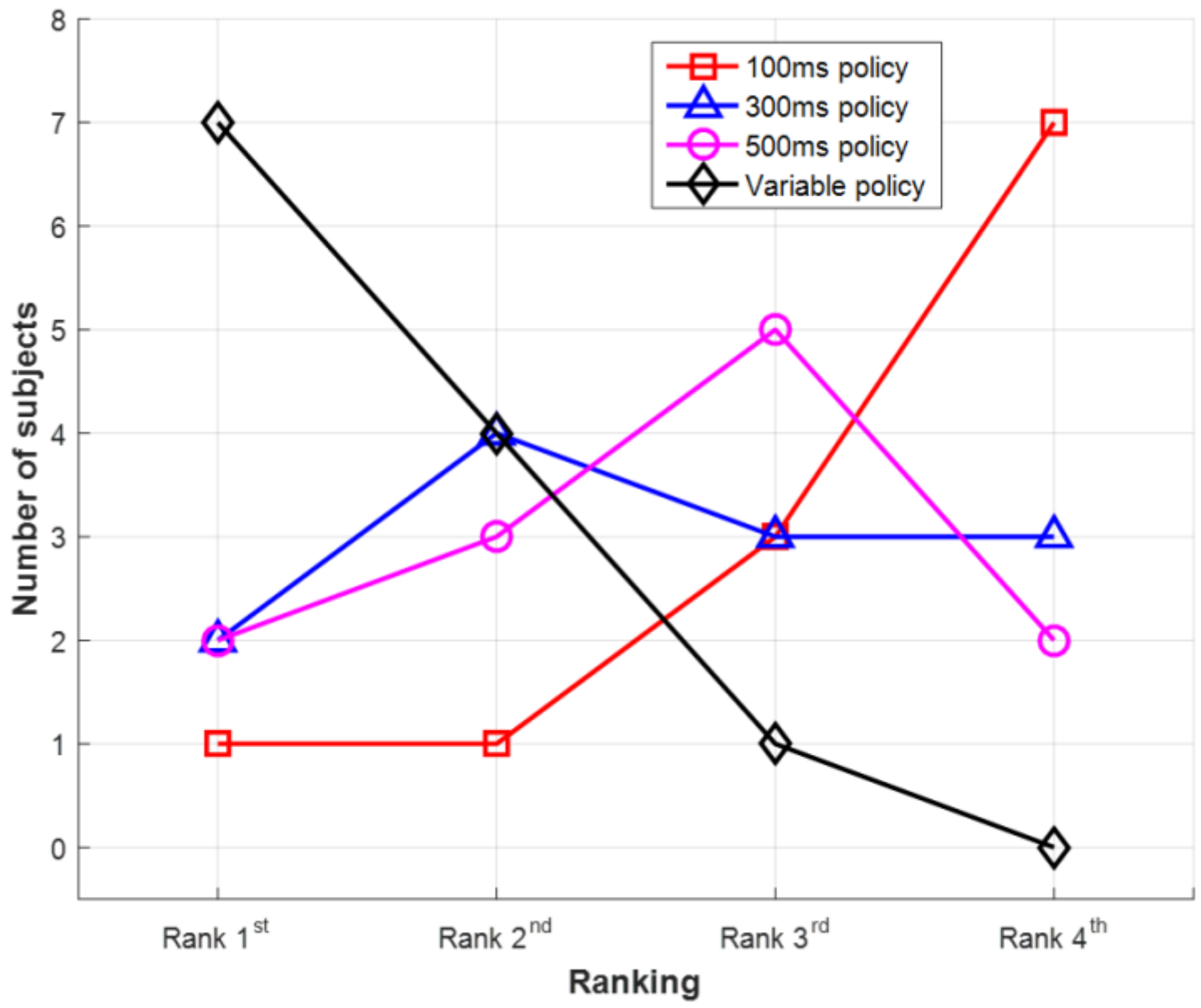

*Figure 13*. Histograms of the overall ranks given to the different dwell time policies by users in Experiment II.

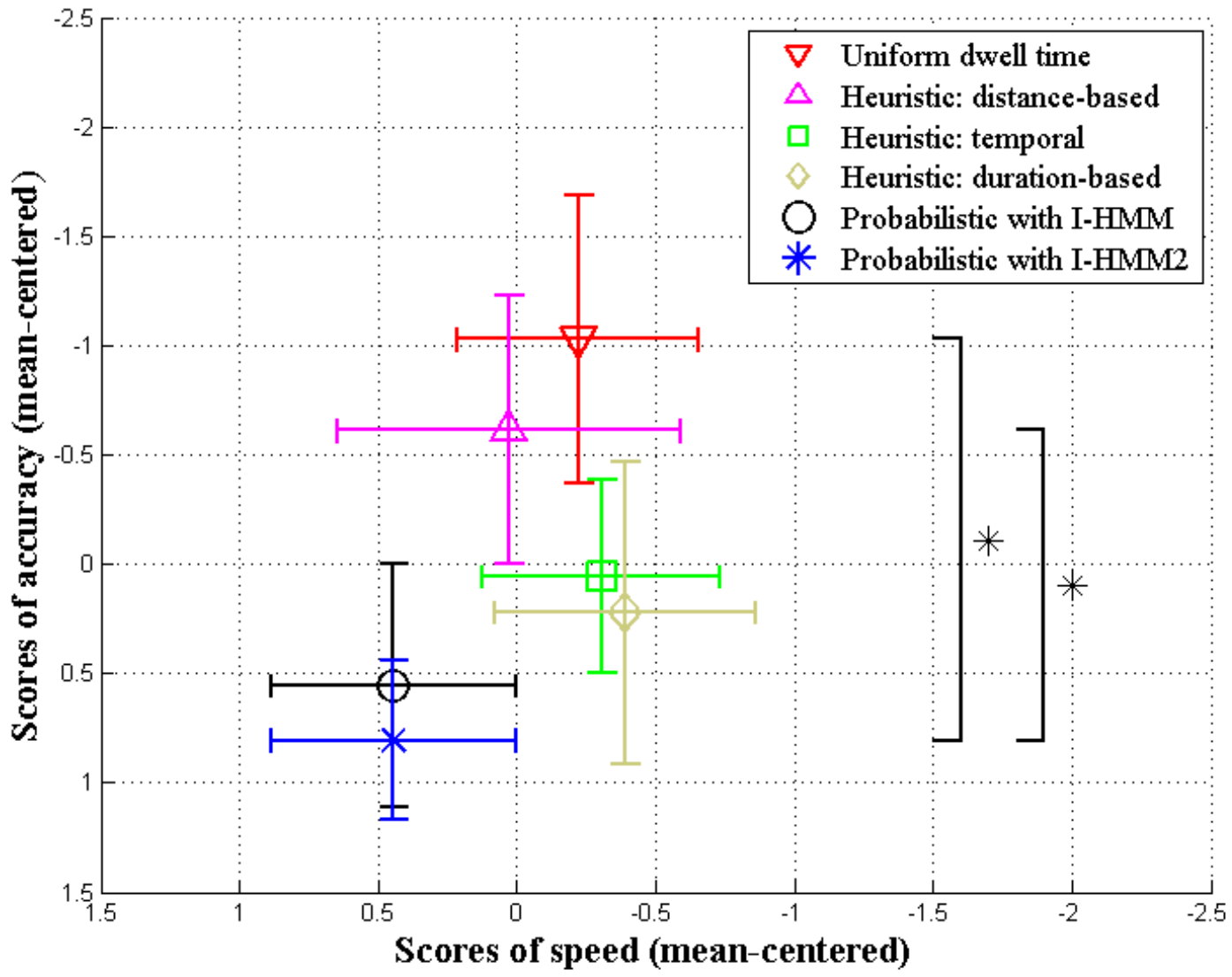

*Figure 14*. The subjective accuracy and speed scores collected in Experiment IV. The scores of each individual subject are mean-centered by subtracting the mean scores of the subject. The error bars represent 95% confidence intervals. The square brackets with asterisks indicate differences with the best probabilistic dwell time policy that are statistically significant ($p < .05$).

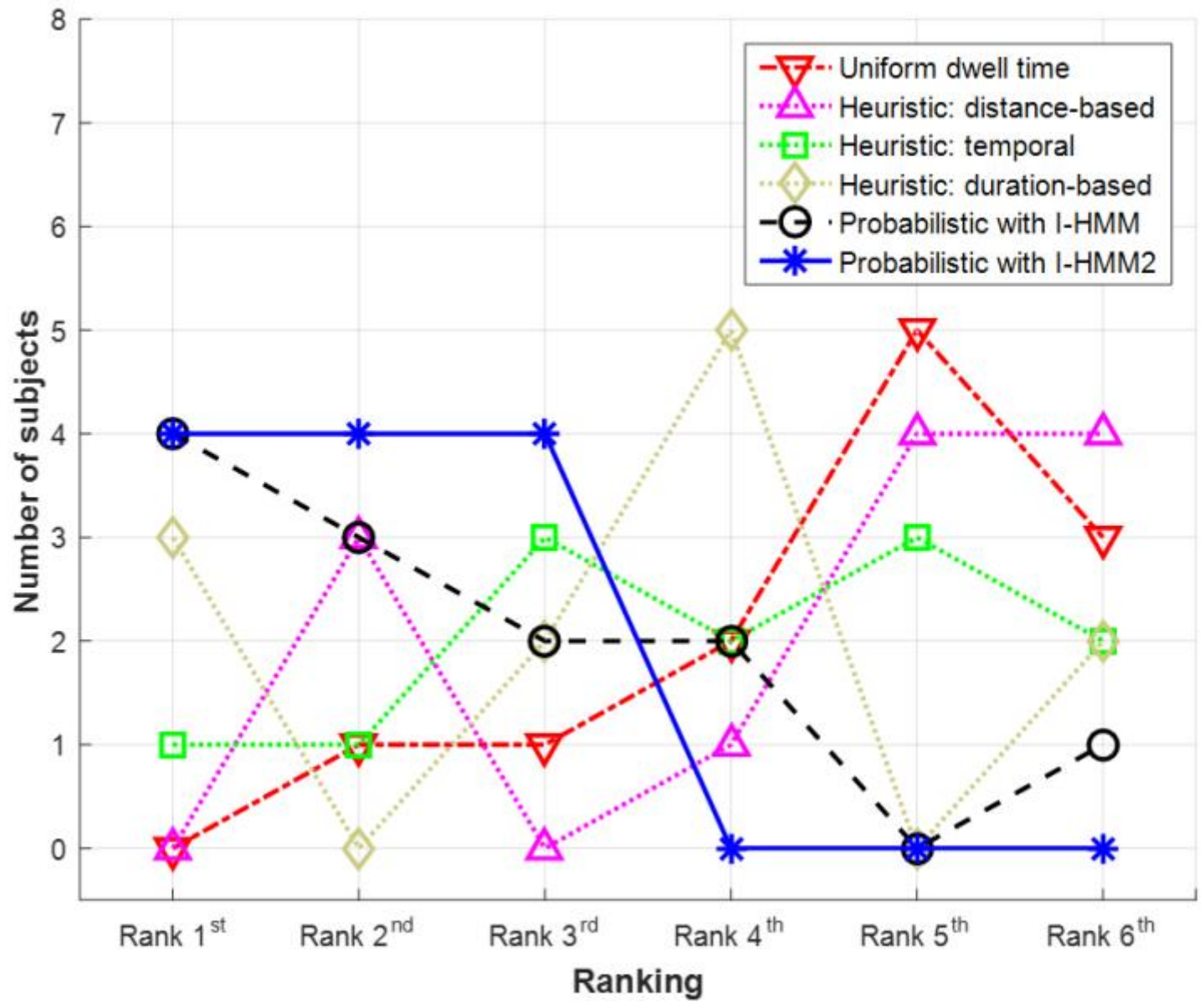

*Figure 15*. Histograms of the overall ranks given to the different dwell time policies by users in Experiment IV.